\documentclass[sigconf]{acmart}

\setcopyright{acmlicensed}
\acmYear{2017}
\copyrightyear{2017}
\acmPrice{15.00}

\usepackage{courier}            
\usepackage[scaled]{helvet} 
\usepackage{url}                  
\usepackage{listings}          
\usepackage{enumitem}      

\usepackage[shortcuts]{extdash} 

\usepackage[normalem]{ulem}
\usepackage{amsmath}
\usepackage{calc}
\usepackage{algorithm}
\usepackage[noend]{algpseudocode}
\usepackage[T1]{fontenc}
\usepackage{listings}
\usepackage{color}
\usepackage{xspace}
\usepackage{array}
\usepackage{multirow}
\usepackage{varwidth}
\usepackage{tabularx}  
\usepackage{ragged2e}  
\usepackage{xcolor,colortbl}
\usepackage{graphicx}
\usepackage{mathpartir}

\usepackage{booktabs}
\usepackage{subfig}

\newcommand{\doi}[1]{doi:~\href{http://dx.doi.org/#1}{\Hurl{#1}}}   

\newcommand{\cutthree}[1]{}

\newcommand\crule[3][black]{\textcolor{#1}{\rule{#2}{#3}}}

\makeatletter

\renewcommand\paragraph{\@startsection{paragraph}{4}{\z@}%
  {-.5\baselineskip \@plus -2\p@ \@minus -.2\p@}%
  {-3.5\p@}%
  {\bfseries\@parfont}}

\makeatletter

\newcolumntype{C}[1]{>{\centering}m{#1}}

\newcommand{\TSAdded}[1]{#1}

\newcommand{\myfig}{Fig.~}
\newcommand{\myfiglong}{Figure~}
\newcommand{\myfigs}{Figs.~}
\newcommand{\mytab}{Tab.~}
\newcommand{\mytablong}{Table~}
\newcommand{\mysec}{Sec.~}

\newcommand{\nvidia}{Nvidia\xspace}

\definecolor{Gray3}{gray}{0.9}
\definecolor{Gray2}{gray}{0.75}
\definecolor{Gray1}{gray}{0.6}

\newcommand{\transmit}{\mathsf{transmit}}

\newcommand{\code}[1]{\lstset{basicstyle=\tt}\lstinline!#1!\lstset{basicstyle=\scriptsize\tt}}

\newcommand{\offerfork}{\mathsf{request\_fork}}
\newcommand{\offerkill}{\mathsf{offer\_kill}}
\newcommand{\globalbarrier}{\mathsf{global\_barrier}}
\newcommand{\resizingglobalbarrier}{\mathsf{resizing\_global\_barrier}}
\newcommand{\getgroupid}{\mathsf{get\_group\_id}}
\newcommand{\getnumgroups}{\mathsf{get\_num\_groups}}
\newcommand{\getlocalid}{\mathsf{get\_local\_id}}
\newcommand{\getglobalid}{\mathsf{get\_global\_id}}
\newcommand{\getlocalsize}{\mathsf{get\_local\_size}}
\newcommand{\getglobalsize}{\mathsf{get\_global\_size}}

\newcommand{\keyword}[1]{\mathsf{#1}}

\usepackage{flushend}
\begin{document}

\title{Cooperative Kernels: GPU Multitasking for Blocking Algorithms (Extended Version)}

\author{Tyler Sorensen}
\affiliation{%
  \institution{Imperial College London}
  \city{London}
  \country{UK}}
\email{t.sorensen15@imperial.ac.uk}

\author{Hugues Evrard}
\affiliation{%
  \institution{Imperial College London}
  \city{London}
  \country{UK}}
\email{h.evrard@imperial.ac.uk}

\author{Alastair F. Donaldson}
\affiliation{%
  \institution{Imperial College London}
  \city{London}
  \country{UK}}
\email{alastair.donaldson@imperial.ac.uk}

\begin{abstract}
There is growing interest in accelerating irregular data-parallel
algorithms on GPUs.  These algorithms are typically \emph{blocking},
so they require fair scheduling.  But GPU programming models
(e.g.\ OpenCL) do not mandate fair scheduling, and GPU schedulers are
unfair in practice.  Current approaches avoid this issue by exploiting
scheduling quirks of today's GPUs in a manner that does not allow the
GPU to be shared with other workloads (such as graphics rendering
tasks).  We propose \emph{cooperative kernels}, an extension to the
traditional GPU programming model geared towards writing blocking
algorithms.  Workgroups of a cooperative kernel \emph{are} fairly
scheduled, and multitasking is supported via a small set of language
extensions through which the kernel and scheduler cooperate.  We
describe a prototype implementation of a cooperative kernel framework
implemented in OpenCL 2.0 and evaluate our approach by porting a set
of blocking GPU applications to cooperative kernels and examining
their performance under multitasking.  Our prototype exploits no
vendor-specific hardware, driver or compiler support, thus our results
provide a lower-bound on the efficiency with which cooperative kernels
can be implemented in practice.

\end{abstract}

\ccsdesc[500]{Software and its engineering~Multiprocessing / multiprogramming / multitasking}
\ccsdesc[100]{Software and its engineering~Semantics}
\ccsdesc[300]{Computing methodologies~Graphics processors}

\keywords{GPU, cooperative multitasking, irregular parallelism}

%
%




\maketitle

\newcommand{\myparagraph}[1]{\paragraph{#1.}}

\section{Introduction}\label{sec:intro}

\myparagraph{The Needs of Irregular Data-parallel Algorithms}
Many interesting data-parallel algorithms are \emph{irregular}: the
amount of work to be processed is unknown ahead of time and may change dynamically in a workload-dependent manner.
There is growing interest in
accelerating such algorithms on
GPUs~\cite{owens-persistent,DBLP:conf/ipps/KaleemVPHP16,DBLP:conf/ipps/DavidsonBGO14,DBLP:conf/hipc/HarishN07,DBLP:journals/topc/MerrillGG15,DBLP:conf/egh/VineetHPN09,DBLP:conf/ppopp/NobariCKB12,DBLP:conf/hpcc/SolomonTT10a,DBLP:conf/popl/PrabhuRMH11,DBLP:conf/ppopp/Mendez-LojoBP12,DBLP:conf/oopsla/PaiP16,DBLP:conf/oopsla/SorensenDBGR16,DBLP:conf/egh/CedermanT08,TPO10,BNP12,Pannotia}.
Irregular algorithms usually require \emph{blocking synchronization}
between workgroups, e.g.\ many graph algorithms use a level-by-level
strategy, with a global barrier between levels; work
stealing algorithms require each workgroup to maintain a queue,
typically mutex-protected, to enable stealing by other
workgroups.

To avoid starvation, a blocking concurrent algorithm requires
\emph{fair} scheduling of workgroups.  For
example, if one workgroup holds a mutex, an unfair scheduler may cause
another workgroup to spin-wait forever for the mutex to be
released.  Similarly, an unfair scheduler can cause a workgroup to spin-wait
indefinitely at a global barrier so that other workgroups do not reach the barrier.

\myparagraph{A Degree of Fairness: Occupancy-bound Execution} The current GPU programming
models---OpenCL~\cite{opencl2Spec}, CUDA~\cite{cuda-75} and
HSA~\cite{HSAprogramming11}, specify almost no guarantees regarding
scheduling of workgroups, and current GPU schedulers are unfair in
practice.  Roughly speaking, each workgroup executing a GPU kernel is
mapped to a hardware \emph{compute unit}.\footnote{In practice,
  depending on the kernel, multiple workgroups might map to the same
  compute unit; we ignore this in our current discussion.}
The simplest way for a GPU driver to handle more workgroups being
launched than there are compute units is via an \emph{occupancy-bound}
execution
model~\cite{owens-persistent,DBLP:conf/oopsla/SorensenDBGR16} where,
once a workgroup has commenced execution on a compute unit (it has
become \emph{occupant}), the workgroup has exclusive access to the
compute unit until it finishes execution.
Experiments suggest that this model
is widely employed by today's
GPUs~\cite{owens-persistent,DBLP:conf/oopsla/SorensenDBGR16,DBLP:conf/oopsla/PaiP16,BNP12}.

\begin{figure}[t]
\centering
\includegraphics[width=\columnwidth]{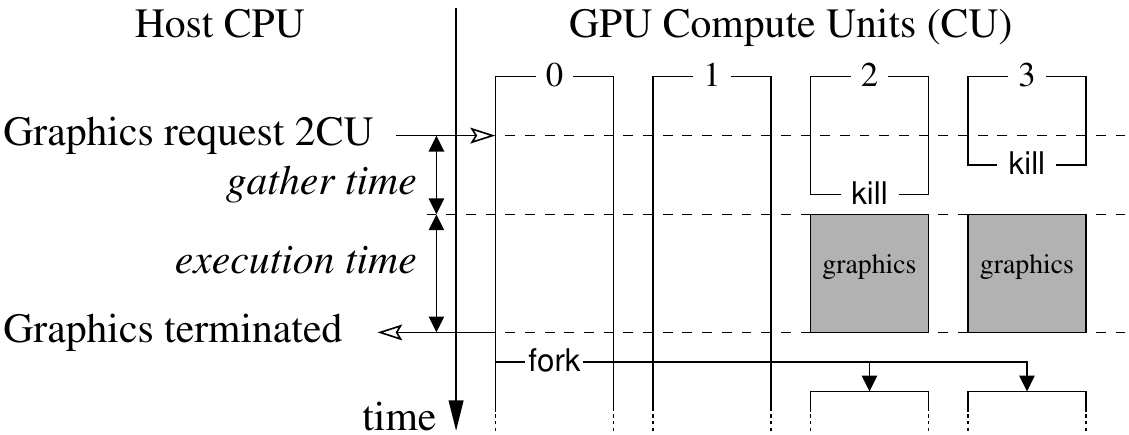}
\caption{Cooperative kernels can flexibly resize to let other tasks,
e.g.\ graphics, run concurrently}
\label{fig:overview}
\end{figure}

The occupancy-bound execution model does not guarantee fair scheduling
between workgroups: if all compute units are occupied then a
not-yet-occupant workgroup will not be scheduled until some occupant
workgroup completes execution.  Yet the execution model \emph{does}
provide fair scheduling between \emph{occupant} workgroups, which are
bound to separate compute units that operate in parallel.  Current GPU
implementations of blocking algorithms assume the occupancy-bound
execution model, which they exploit by launching no more workgroups
than there are available compute units~\cite{owens-persistent}.

\myparagraph{Resistance to Occupancy-bound Execution}
Despite its practical prevalence, none of the current GPU programming
models actually mandate occupancy-bound execution.  Further, there are
reasons why this model is undesirable.
First, the execution model does not enable
multitasking, since a workgroup effectively \emph{owns} a compute
unit until the workgroup has completed execution.  The GPU cannot be used meanwhile for other
tasks (e.g.\ rendering).
Second, \emph{energy throttling} is an important concern for
battery-powered
devices~\cite{DBLP:journals/comsur/Vallina-RodriguezC13}.  In the
future, it will be desirable for a mobile GPU driver to power down
some compute units, suspending execution of associated occupant
workgroups, if the battery level is low.

Our assessment, informed by discussions with a number of industrial
practitioners who have been involved in the OpenCL and/or HSA
standardisation efforts
(including~\cite{PersonalCommunicationRichards,PersonalCommunicationHowes}),
is that GPU vendors (1) will not commit to the occupancy-bound
execution model they currently implement, for the above reasons, yet
(2) will not guarantee fair scheduling \TSAdded{using preemption. This
  is due to the high runtime cost of preempting workgroups, which
  requires managing thread local state (e.g. registers, program
  location) for all workgroup threads (up to 1024 on Nvidia
  GPUs), as well as \emph{shared memory}, the workgroup local cache
  (up to 64 KB on Nvidia GPUs)}.  Vendors instead wish to retain the
essence of the simple occupancy-bound model, supporting preemption
only in key special cases.

For example, preemption is supported by Nvidia's Pascal architecture~\cite{PascalWhitepaper}, but on a GTX Titan X (Pascal)
we still observe starvation: a global barrier
executes successfully with 56 workgroups, but deadlocks with 57
workgroups, indicating unfair scheduling.

\myparagraph{Our Proposal: Cooperative Kernels}
To summarise: blocking algorithms
demand fair scheduling,
but for good reasons
GPU vendors will not commit to the guarantees of the
occupancy-bound execution model.
We propose \emph{cooperative kernels}, an extension to the GPU
programming model that aims to resolve this impasse.

A kernel
that requires fair scheduling is identified as \emph{cooperative}, and written using two additional
language primitives, $\offerkill$ and $\offerfork$, placed by the programmer.
Where the cooperative kernel could proceed with fewer
workgroups, a workgroup can execute $\offerkill$, offering to
sacrifice itself to the scheduler.  This indicates that the workgroup
would ideally continue executing, but that the
scheduler may preempt the workgroup; the cooperative kernel
must be prepared to deal with either scenario.
Where the cooperative kernel could use additional resources, a workgroup can execute
$\offerfork$ to indicate that the
kernel is prepared to proceed with the existing set of workgroups, but
is able to benefit from one or more additional workgroups
commencing execution directly after the $\offerfork$ program point.

The use of $\offerfork$ and $\offerkill$ creates a contract between
the scheduler and the cooperative kernel.  Functionally, the scheduler
must guarantee that the workgroups executing a cooperative kernel are
fairly scheduled, while the cooperative kernel must be robust to
workgroups leaving and joining the computation in response to
$\offerkill$ and $\offerfork$.  Non-functionally, a cooperative kernel
must ensure that $\offerkill$ is executed frequently enough such that
the scheduler can accommodate soft-real time constraints,
e.g.\ allowing a smooth frame-rate for graphics.
In return, the scheduler should allow the cooperative kernel to
utilise hardware resources where possible, killing workgroups only
when demanded by other tasks, and forking additional workgroups when
possible.

\TSAdded{Cooperative kernels allow for \emph{cooperative multitasking}
  (see \mysec{\ref{sec:relatedwork}}), used
  historically when preemption was not available or too costly. Our approach avoids the cost of arbitrary
  preemption as the state of a workgroup killed via $\offerkill$ does
  not have to be saved. Previous cooperative multitasking systems have
  provided \emph{yield} semantics, where a processing unit would
  temporarily give up its hardware resource. We deviate from this
  design as, in the case of a global barrier, adopting yield would
  force the cooperative kernel to block \emph{completely} when a single workgroup yields, stalling the kernel until the given
  workgroup resumes.  Instead, our $\offerkill$ allows a kernel to
  make progress with a smaller number of workgroups, with workgroups
  potentially joining again later via $\offerfork$.}

\myfiglong{\ref{fig:overview}} illustrates sharing of GPU compute units between a cooperative kernel and a
graphics task. Workgroups 2 and 3 of the cooperative kernel
are killed at an $\offerkill$ to make room for a graphics
task. The workgroups are subsequently restored to the cooperative kernel when workgroup 0 calls $\offerfork$. The \emph{gather time} is the time
between resources being requested and the application surrendering them via $\offerkill$. To satisfy soft-real time
constraints, this time should be low; our experimental study
(\mysec{\ref{sec:responsiveness}}) shows that, in practice, the
gather-time for our applications is acceptable for a range of graphics
workloads.

The cooperative kernels model has several appealing
properties:

\begin{enumerate}[leftmargin=*]

\item By providing fair scheduling between workgroups, cooperative
  kernels meet the needs of blocking algorithms, including irregular
  data-parallel algorithms.

\item The model has no impact on the development of regular
  (non-cooperative) compute and graphics kernels.

\item The model is backwards-compatible: $\offerkill$ and $\offerfork$ may be ignored, and a cooperative kernel will behave
  exactly as a regular kernel does on current GPUs.

\item Cooperative kernels can be implemented over the occupancy\-/bound
  execution model provided by current GPUs: our prototype implementation uses no special hardware/driver support.

\item If hardware support for preemption \emph{is} available, it can be leveraged to implement cooperative kernels efficiently, and cooperative kernels can avoid unnecessary preemptions by allowing the programmer to communicate ``smart'' preemption points.

\end{enumerate}

Placing the primitives manually is straightforward for the representative set of
GPU-accelerated irregular algorithms we have ported so far.  Our experiments show that the model can enable efficient multitasking of cooperative and non-cooperative tasks.

In summary, our main contributions are:
\emph{cooperative kernels}, an extended GPU programming model that supports the scheduling requirements of blocking algorithms (\mysec\ref{sec:cooperativekernels}); a \emph{prototype implementation} of cooperative
  kernels on top of OpenCL 2.0
  (\mysec\ref{sec:implementation}); and \emph{experiments} assessing the overhead and responsiveness of the cooperative kernels approach over a set of irregular algorithms \cutthree{across three GPUs} (\mysec\ref{sec:experiments}), including a best-effort comparison with the efficiency afforded by hardware-supported preemption available on Nvidia GPUs.

We begin by providing background on OpenCL via two motivating examples (\mysec\ref{sec:background}).  At the end we discuss related work (\mysec\ref{sec:relatedwork}) and avenues for future work (\mysec\ref{sec:conclusion}).

\section{Background and Examples}\label{sec:background}

We outline the OpenCL programming model on which we
base cooperative kernels (\mysec\ref{sec:opencl}), and illustrate
OpenCL and the scheduling requirements of irregular algorithms using two examples: a work stealing queue and frontier-based graph traversal
(\mysec\ref{sec:openclexamples}).

\subsection{OpenCL Background}\label{sec:opencl}


An OpenCL program is divided into \emph{host} and \emph{device}
components.  A host application runs on the CPU and launches one or
more \emph{kernels} that run on accelerator devices---GPUs in the
context of this paper.  A kernel is written in OpenCL C, based on C99.
All threads executing a kernel start at the same entry function with
identical arguments.  A thread can call $\getglobalid$
to obtain a unique id, to access distinct data or follow different control flow paths.

The threads of a kernel are divided into \emph{workgroups}.
Functions
$\getlocalid$ and $\getgroupid$ return a thread's local id within
its workgroup and the workgroup id.
%
%
The number
of threads per workgroup and number of workgroups are obtained via
$\getlocalsize$ and $\getnumgroups$.
%
Execution of the threads in a workgroup can be synchronised via a
workgroup barrier.
A \emph{global} barrier (synchronising all
threads of a kernel) is \emph{not} provided as a primitive.

\myparagraph{Memory Spaces and Memory Model} A kernel has access to
four memory spaces.  \emph{Shared virtual memory} (SVM) is accessible
to all device threads and the host concurrently.  \emph{Global} memory is
shared among all device threads.  Each workgroup has a
portion of \emph{local} memory for fast intra-workgroup communication.
Every thread has a portion of very fast \emph{private} memory for
function-local variables.

Fine-grained
communication within a workgroup, as well as inter-workgroup
communication and communication with the host while the kernel is
running, is enabled by a set of atomic data types and operations.  In
particular, fine-grained host/device communication is via atomic
operations on SVM.

\myparagraph{Execution Model}
OpenCL~\cite[p.\ 31]{opencl2Spec} and CUDA~\cite{cuda-75} specifically make no guarantees about fair scheduling between
workgroups executing the same kernel.
%
HSA provides limited, one-way guarantees,
stating~\cite[p. 46]{HSAprogramming11}: \emph{``Work-group A can wait
  for values written by work-group B without deadlock provided ... (if) A
  comes after B in work-group flattened ID order''}. This is not sufficient to support blocking algorithms that use
mutexes and inter-workgroup barriers, both of which require \emph{symmetric} communication between
threads.

\subsection{Motivating Examples}\label{sec:openclexamples}


\begin{figure}[t]


\begin{lstlisting}
(*@\label{line:wksteal:kernelfunc}@*)kernel work_stealing(global Task * queues) {
(*@\label{line:wksteal:getgroupid}@*)  int queue_id = get_group_id();
(*@\label{line:wksteal:mainloop}@*)  while (more_work(queues)) {
(*@\label{line:wksteal:poporsteal}@*)    Task * t = pop_or_steal(queues, queue_id);
    if (t)
(*@\label{line:wksteal:processtask}@*)      process_task(t, queues, queue_id);
  }
}
\end{lstlisting}
\caption{An excerpt of a work stealing algorithm in OpenCL}\label{fig:workstealing}
\end{figure}

\myparagraph{Work Stealing}
Work stealing enables dynamic balancing of tasks across
processing units. It is useful when the number of tasks to be
processed is dynamic, due to one task creating an arbitrary number of
new tasks.  Work stealing has been explored in the context of
GPUs~\cite{DBLP:conf/egh/CedermanT08,TPO10}. Each workgroup has a
queue from which it obtains tasks to process, and to which
it stores new tasks. If its queue is empty, a workgroup
tries to \emph{steal} a task from another queue.

\myfiglong\ref{fig:workstealing} illustrates a work stealing
kernel. Each thread receives a pointer to the task queues, in global
memory, initialized by the host to contain initial tasks. A thread
uses its workgroup id (line~\ref{line:wksteal:getgroupid}) as a queue
id to access the relevant task queue. The $\mathsf{pop\_or\_steal}$
function (line~\ref{line:wksteal:poporsteal}) pops a task from the
workgroup's queue or tries to steal a task from other queues. Although
not depicted here, concurrent accesses to queues inside
$\mathsf{more\_work}$ and $\mathsf{pop\_or\_steal}$ are guarded by a
mutex per queue, implemented using atomic compare and swap operations
on global memory.

If a task is obtained, then the workgroup processes it
(line~\ref{line:wksteal:processtask}), which may lead to new tasks
being created and pushed to the workgroup's queue. The kernel presents
two opportunities for spin-waiting: spinning to obtain a mutex, and
spinning in the main kernel loop to obtain a task. \TSAdded{Without fair
scheduling, threads waiting for the mutex might spin indefinitely,
causing the application to hang.}

%
%

\begin{figure}[t]

\begin{lstlisting}
kernel graph_app(global graph * g,
       global nodes * n0, global nodes * n1) {
  int level = 0;
  global nodes * in_nodes = n0;
  global nodes * out_nodes = n1;
  int tid = get_global_id();
  int stride = get_global_size();
(*@\label{line:graph:iterate}@*)  while(in_nodes.size > 0) {
    for (int i = tid; i < in_nodes.size; i += stride)
      process_node(g, in_nodes[i], out_nodes, level);
(*@\label{line:graph:swap}@*)    swap(&in_nodes, &out_nodes);
(*@\label{line:graph:gb1}@*)    global_barrier();
(*@\label{line:graph:reset}@*)    reset(out_nodes);
    level++;
(*@\label{line:graph:gb2}@*)    global_barrier();
  }
}
\end{lstlisting}
\caption{An OpenCL graph traversal algorithm}\label{fig:graphsearch}
\end{figure}

\myparagraph{Graph Traversal} \myfiglong\ref{fig:graphsearch} illustrates a frontier-based graph traversal algorithm; such algorithms have
been shown to execute efficiently on GPUs~\cite{BNP12,DBLP:conf/oopsla/PaiP16}.
The kernel is
given three arguments in global memory: a graph structure, and two
arrays of graph nodes. Initially, $\keyword{n0}$ contains the
starting nodes to process. Private variable $\keyword{level}$ records the current frontier level, and $\keyword{in\_nodes}$ and $\keyword{out\_nodes}$ point to
distinct arrays recording the nodes to be processed during the current and next frontier, respectively.

The application iterates as long as the current frontier contains
nodes to process (line~\ref{line:graph:iterate}). At each frontier,
the nodes to be processed are evenly distributed between
threads through \emph{stride} based processing.
In this case, the stride is the total number of threads, obtained via
$\getglobalsize$.
A thread calls $\keyword{process\_node}$ to process a node given the current level, with nodes to be processed during the next frontier being pushed to $\keyword{out\_nodes}$. After processing the frontier, the threads swap their
node array pointers (line~\ref{line:graph:swap}).

At this point, the GPU threads must wait for all other threads to
finish processing the frontier. To achieve
this, we use a global barrier construct
(line~\ref{line:graph:gb1}). After all threads reach this point, the
output node array is reset (line~\ref{line:graph:reset}) and the level
is incremented. The threads use another global barrier to wait until the output node is
reset (line~\ref{line:graph:gb2}), after which they continue to the next frontier.

The global barrier used in this application is not provided as a GPU
primitive, though previous works have shown that such a global barrier
can be implemented~\cite{XF10,DBLP:conf/oopsla/SorensenDBGR16}, based
on CPU barrier designs~\cite[ch. 17]{HS08}.  These barriers employ
spinning to ensure each thread waits at the barrier until all threads have
arrived, thus fair scheduling between workgroups is required for the
barrier to operate correctly. \TSAdded{Without fair scheduling, the
  barrier threads may wait indefinitely at the barrier, causing the
  application to hang.}

The mutexes and barriers used by these two examples appear to run
reliably on current GPUs for kernels that are executed with no more
workgroups than there are compute units.  This is due to the fairness
of the occupancy-bound execution model that current GPUs have been
shown, experimentally, to provide.  But, as discussed in
\mysec\ref{sec:intro}, this model is not endorsed by language
standards or vendor implementations, and may not be respected
in the future.

In \mysec\ref{sec:programmingguidelines} we show how the work stealing
and graph traversal examples of \myfigs\ref{fig:workstealing} and~\ref{fig:graphsearch} can be
updated to use our cooperative kernels programming model to resolve
the scheduling issue.

\section{Cooperative Kernels}\label{sec:cooperativekernels}

We present our cooperative kernels programming model as an extension
to OpenCL.
We describe the semantics of the model
(\mysec\ref{sec:semantics}), presenting a more formal operational semantics in Appendix~\ref{appendix:semantics}
and discussing possible alternative semantic choices in Appendix~\ref{appendix:semanticalternatives}, use our motivating examples to discuss
programmability (\mysec\ref{sec:programmingguidelines}) and outline
important nonfunctional properties that the model requires to work
successfully (\mysec\ref{sec:nonfunctional}).
%

\subsection{Semantics of Cooperative Kernels}\label{sec:semantics}

As with a regular OpenCL kernel, a cooperative kernel is launched by
the host application, passing parameters to the kernel and specifying
a desired number of threads and workgroups. Unlike in a regular
kernel, the parameters to a cooperative kernel are immutable (though
pointer parameters can refer to mutable data).

Cooperative kernels are written using the following
extensions: $\transmit$, a qualifier on the variables of a
thread; $\offerkill$ and $\offerfork$, the key functions that enable
cooperative scheduling; and $\globalbarrier$ and $\resizingglobalbarrier$
primitives for inter-workgroup synchronisation.

\myparagraph{Transmitted Variables}
A variable declared in the root scope of the cooperative kernel can
optionally be annotated with a new $\transmit$ qualifier.  Annotating
a variable $v$ with $\transmit$ means that when a workgroup
uses $\offerfork$ to spawn new
workgroups, the workgroup should transmit its
current value for $v$ to the threads of the new workgroups.
We detail the semantics for this when we
describe $\offerfork$ below.

\myparagraph{Active Workgroups}
If the host application launches a cooperative kernel requesting $N$
workgroups, this indicates that the kernel should be executed with a
\emph{maximum} of $N$ workgroups, and that as many workgroups as possible, up
to this limit, are desired.  However, the scheduler may initially
schedule fewer than $N$ workgroups, and as explained below the number
of workgroups that execute the cooperative kernel can change during
the lifetime of the kernel.

The number of \emph{active workgroups}---workgroups executing the
kernel---is denoted $M$.  Active workgroups have consecutive ids in
the range $[0, M-1]$.  Initially, at least one workgroup is active; if
necessary the scheduler must postpone the kernel until some compute
unit becomes available. For example, in \myfig{\ref{fig:overview}: at the
  beginning of the execution $M = 4$; while the graphics task is
  executing $M = 2$; after the fork $M = 4$ again.

When executed by a cooperative
kernel, $\getnumgroups$ returns $M$, the \emph{current} number of
active workgroups.  This is in contrast to $\getnumgroups$ for
regular kernels, which returns the fixed number of workgroups that execute the kernel (see \mysec\ref{sec:opencl}).

Fair scheduling \emph{is} guaranteed between active workgroups;
i.e.\ if some thread in an active workgroup is enabled, then
eventually this thread is guaranteed to execute an instruction.


\myparagraph{Semantics for $\offerkill$}
The $\offerkill$ primitive allows the cooperative kernel to return
compute units to the scheduler by offering to sacrifice workgroups.
The idea is as follows: allowing the scheduler to arbitrarily and abruptly terminate execution
of workgroups might be drastic, yet the kernel
may contain specific program points at which a workgroup could
\emph{gracefully} leave the computation.

Similar to the OpenCL workgroup $\mathsf{barrier}$ primitive,
$\offerkill$, is a workgroup-level function---it must be encountered
uniformly by all threads in a workgroup.

Suppose a workgroup with id $m$ executes $\offerkill$.  If the
workgroup has the largest id among active workgroups then it
can be killed by the scheduler, except that workgroup 0 can never be
killed (to avoid early termination of the kernel).  More formally, if $m < M-1$ or $M=1$ then $\offerkill$ is a
no-op.  If instead $M > 1$ and $m = M-1$, the scheduler can choose to
ignore the offer, so that $\offerkill$ executes as a no-op, or accept
the offer, so that execution of the workgroup ceases and the number of
active workgroups $M$ is atomically decremented by one.
\myfiglong{\ref{fig:overview}} illustrates this, showing that
workgroup $3$ is killed before workgroup $2$.

\myparagraph{Semantics for $\offerfork$}
Recall that a desired limit of $N$ workgroups was specified when the
cooperative kernel was launched, but that the number of active
workgroups, $M$, may be smaller than $N$, either because (due to
competing workloads) the scheduler did not provide $N$ workgroups
initially, or because the kernel has given up some workgroups via
$\offerkill$ calls.  Through the $\offerfork$ primitive (also a
workgroup-level function), the kernel and scheduler can collaborate to
allow new workgroups to join the computation at an appropriate point
and with appropriate state.

Suppose a workgroup with id $m\leq M$ executes $\offerfork$.  Then the
following occurs: an integer $k \in [0, N-M]$ is chosen by the
scheduler; $k$ new workgroups are spawned with consecutive ids in the
range $[M, M+k-1]$; the active workgroup count $M$ is atomically
incremented by $k$.

The $k$ new workgroups commence execution at the program point
immediately following the $\offerfork$ call.  The variables that
describe the state of a thread are all uninitialised for the threads
in the new workgroups; reading from these variables without first
initialising them is an undefined behaviour.  There are two exceptions
to this: (1) because the parameters to a cooperative kernel are
immutable, the new threads have access to these parameters as part of
their local state and can safely read from them; (2) for each variable
$v$ annotated with $\transmit$, every new thread's copy of $v$ is
initialised to the value that thread 0 in workgroup $m$ held for $v$
at the point of the $\offerfork$ call.
In effect, thread 0 of the forking workgroup \emph{transmits} the relevant
portion of its local state to the threads of the forked workgroups.

\myfiglong{\ref{fig:overview}} illustrates the behaviour of
$\offerfork$. After the graphics task finishes executing, workgroup
$0$ calls $\offerfork$, spawning the two new workgroups with ids $2$
and $3$. Workgroups $2$ and $3$ join the computation where workgroup
$0$ called $\offerfork$.
%
%

Notice that $k=0$ is always a valid choice for the number of
workgroups to be spawned by $\offerfork$, and is guaranteed if $M$ is
equal to the workgroup limit $N$.

\myparagraph{Global Barriers}
Because workgroups of a cooperative kernel are fairly scheduled, a
global barrier primitive can be provided.  We specify two variants: $\globalbarrier$
and $\resizingglobalbarrier$.

Our $\globalbarrier$ primitive is a kernel-level function: if it
appears in conditional code then it must be reached by \emph{all}
threads executing the cooperative kernel.  On reaching a
$\globalbarrier$, a thread waits until all threads have arrived at
the barrier.  Once all threads have arrived, the threads may proceed
past the barrier with the guarantee that all global memory accesses
issued before the barrier have completed.  The $\globalbarrier$
primitive can be implemented by adapting an inter-workgroup barrier
design, e.g.~\cite{XF10}, to take account of a growing and shrinking number of workgroups, and the atomic operations provided by
the OpenCL 2.0 memory model enable a memory-safe
implementation~\cite{DBLP:conf/oopsla/SorensenDBGR16}.

The $\resizingglobalbarrier$ primitive is also a kernel-level
function.  It is identical to $\globalbarrier$, except that it caters
for cooperation with the scheduler: by issuing a
$\resizingglobalbarrier$ the programmer indicates that the cooperative
kernel is prepared to proceed after the barrier with more or fewer workgroups.

When all threads have reached $\resizingglobalbarrier$,
the number of active workgroups, $M$, is atomically set to a new value, $M'$ say, with $0 < M' \leq N$.
If $M' = M$ then the active workgroups remain unchanged.  If $M' < M$, workgroups $[M', M-1]$ are
killed.  If $M' > M$ then $M'-M$ new workgroups join the computation after the barrier,
as if they were forked from workgroup 0.  In particular, the
$\transmit$-annotated local state of thread 0 in workgroup 0 is
transmitted to the threads of the new workgroups.

The semantics of $\resizingglobalbarrier$ can be modelled via calling $\offerfork$ and $\offerkill$,
surrounded and separated by calls to a $\globalbarrier$.
%
%
The enclosing $\globalbarrier$ calls ensure that the change in number
of active workgroups from $M$ to $M'$ occurs entirely within the
resizing barrier, so that $M$ changes atomically from a programmer's perspective.  The middle $\globalbarrier$ ensures that forking occurs
before killing, so that workgroups $[0, \textrm{min}(M, M') - 1]$ are
left intact.

Because $\resizingglobalbarrier$ can be implemented as above, we do
not regard it \emph{conceptually} as a primitive of our model.
However, in \mysec\ref{sec:resizingbarrier} we show how a resizing
barrier can be implemented more efficiently through direct interaction
with the scheduler.

\subsection{Programming with Cooperative Kernels}\label{sec:programmingguidelines}


\myparagraph{A Changing Workgroup Count} Unlike in regular OpenCL, the
value returned by $\getnumgroups$ is not fixed during the lifetime of
a cooperative kernel: it corresponds to the active group count $M$,
which changes as workgroups execute $\offerkill$, and $\offerfork$.
The value returned by $\getglobalsize$ is similarly subject to change.
A cooperative kernel must thus be written in a manner that is robust
to changes in the values returned by these functions.

In general, their volatility means that use of these functions should
be avoided.  However, the situation is more stable if a cooperative
kernel does not call $\offerkill$ and $\offerfork$ directly, so that
only $\resizingglobalbarrier$ can affect the number of active
workgroups.  Then, at any point during execution, the threads of a
kernel are executing between some pair of resizing barrier calls,
which we call a \emph{resizing barrier interval} (considering the
kernel entry and exit points conceptually to be special cases of
resizing barriers).  The active workgroup count is constant within
each resizing barrier interval, so that $\getnumgroups$ and
$\getglobalsize$ return stable values during such intervals.
As we illustrate below for graph traversal, this can be exploited by algorithms that perform strided
data processing.

\myparagraph{Adapting Work Stealing}
In this example there is no state to transmit since a computation is
entirely parameterised by a task, which is retrieved from a queue
located in global memory. With respect to \myfig~\ref{fig:workstealing},
we add $\offerfork$ and $\offerkill$ calls at the start of the main loop
(below line~\ref{line:wksteal:mainloop}) to let a workgroup offer itself
to be killed or forked, respectively, before it processes a task.  Note
that a workgroup may be killed even if its associated task queue is not
empty, since remaining tasks will be stolen by other workgroups. In
addition, since $\offerfork$ may be the entry point of a workgroup, the
queue id must now be computed after it, so we move
line~\ref{line:wksteal:getgroupid} to be placed just before
line~\ref{line:wksteal:poporsteal}. In particular, the queue id cannot
be transmitted since we want a newly spawned workgroup to read its own
queue and not the one of the forking workgroup.


\myparagraph{Adapting Graph Traversal}
\myfiglong\ref{fig:cgraphsearch} shows a cooperative version of the
graph traversal kernel of \myfig\ref{fig:graphsearch} from
\mysec\ref{sec:openclexamples}.  On lines~\ref{line:cgraph:resizing1}
and ~\ref{line:cgraph:resizing2}, we change the original global
barriers into a resizing barriers. Several variables are marked to be
transmitted in the case of workgroups joining at the resizing barriers
(lines~\ref{line:cgraph:transmit1}, \ref{line:cgraph:transmit2} and
\ref{line:cgraph:transmit3}): $\keyword{level}$ must be restored so
that new workgroups know which frontier they are processing;
$\keyword{in\_nodes}$ and $\keyword{out\_nodes}$ must be restored so
that new workgroups know which of the node arrays to use for input and
output. Lastly, the static work distribution of the original kernel is
no longer valid in a cooperative kernel. This is because the stride
(which is based on $M$) may change after each resizing barrier
call. To fix this, we re-distribute the work after each resizing
barrier call by recomputing the thread id and stride
(lines~\ref{line:cgraph:rechunking1} and
\ref{line:cgraph:rechunking2}). This example exploits the fact that
the cooperative kernel does not issue $\offerkill$ nor $\offerfork$
directly: the value of $\keyword{stride}$ obtained from
$\getglobalsize$ at line~\ref{line:cgraph:rechunking2} is stable
until the next resizing barrier at line~\ref{line:cgraph:resizing1}.

\begin{figure}

\begin{lstlisting}
kernel graph_app(global graph *g,
       global nodes *n0, global nodes *n1) {
(*@\label{line:cgraph:transmit1}@*)  transmit int level = 0;
(*@\label{line:cgraph:transmit2}@*)  transmit global nodes *in_nodes = n0;
(*@\label{line:cgraph:transmit3}@*)  transmit global nodes *out_nodes = n1;
  while(in_nodes.size > 0) {
(*@\label{line:cgraph:rechunking1}@*)    int tid = get_global_id();
(*@\label{line:cgraph:rechunking2}@*)    int stride = get_global_size();
    for (int i = tid; i < in_nodes.size; i += stride)
      process_node(g, in_nodes[i], out_nodes, level);
    swap(&in_nodes, &out_nodes);
(*@\label{line:cgraph:resizing1}@*)    resizing_global_barrier();
    reset(out_nodes);
    level++;
(*@\label{line:cgraph:resizing2}@*)    resizing_global_barrier();
  }
}
\end{lstlisting}
\caption{Cooperative version of the graph traversal kernel of \myfig\ref{fig:graphsearch}, using a resizing barrier and $\transmit$ annotations}\label{fig:cgraphsearch}
\end{figure}

\myparagraph{Patterns for Irregular Algorithms}
In \mysec\ref{sec:portingalgorithms} we describe the set of irregular GPU algorithms used
in our experiments, which largely captures the irregular blocking
algorithms that are available as open source GPU kernels.  These all
employ either work stealing or operate on graph data structures, and placing our new constructs follows a common, easy-to-follow pattern in each case.
The work stealing algorithms have a transactional flavour
and require little or no state to be carried between transactions.  The point at which a workgroup is ready to process a new task is a natural place for $\offerkill$ and $\offerfork$, and few or no $\transmit$ annotations are required.
\myfiglong\ref{fig:cgraphsearch} is representative of
most level-by-level graph algorithms.
It is typically the case that on completing a level of
the graph algorithm, the next level could be processed by more or
fewer workgroups, which $\resizingglobalbarrier$
facilitates.  Some level-specific state must be transmitted to new workgroups.


\subsection{Non-Functional Requirements}\label{sec:nonfunctional}

The semantics presented in \mysec\ref{sec:semantics} describe the
behaviours that a developer of a cooperative kernel should be prepared
for.
However, the aim of cooperative kernels is to find a balance that
allows \emph{efficient} execution of algorithms that require fair scheduling, and
\emph{responsive} multitasking, so that the GPU can be shared between
cooperative kernels and other shorter tasks with soft real-time constraints.
To achieve this balance, an implementation of the cooperative
kernels model, and the programmer of a cooperative kernel, must strive
to meet the following non-functional requirements.


The purpose of $\offerkill$ is to let the scheduler destroy a workgroup
in order to schedule higher-priority tasks.  The scheduler relies on the
cooperative kernel to execute $\offerkill$ sufficiently frequently that
soft real-time constraints of other workloads can be met.
Using our work stealing example: a workgroup offers itself to
the scheduler after processing each task.  If tasks are sufficiently
fast to process then the scheduler will have ample opportunities to
de-schedule workgroups.  But if tasks are very time-consuming to
process then it might be necessary to rewrite the algorithm so that
tasks are shorter and more numerous, to achieve a higher rate of calls
to $\offerkill$.
Getting this non-functional requirement right is GPU- and
application-dependent.  In \mysec\ref{sec:sizingnoncoop} we conduct
experiments to understand the response rate that would be required to
co-schedule graphics rendering with a cooperative kernel, maintaining
a smooth frame rate.


Recall that, on launch, the cooperative kernel requests $N$ workgroups.
The scheduler should thus aim to provide $N$ workgroups if other constraints allow it,
by accepting an $\offerkill$ only if a compute unit is required for another
task, and responding positively to $\offerfork$ calls if compute units are available.



\section{Prototype Implementation}\label{sec:implementation}

Our vision is that cooperative kernel support will be integrated
in the runtimes of future GPU implementations of OpenCL, with driver
support for our new primitives.  To experiment with our ideas on
current GPUs, we have developed a prototype that mocks up the required
runtime support via a \emph{megakernel}, and exploits the
occupancy-bound execution model that these GPUs provide to ensure fair
scheduling between workgroups.  We emphasise that an aim of
cooperative kernels is to \emph{avoid} depending on the
occupancy-bound model.  Our prototype exploits this model simply to
allow us to experiment with current GPUs whose proprietary drivers we
cannot change.  We describe the megakernel approach
(\mysec\ref{sec:megakernel}) and detail various aspects of the
scheduler component of our implementation
(\mysec\ref{sec:schedulerimpl}).

\subsection{The Megakernel Mock Up}\label{sec:megakernel}

Instead of multitasking multiple separate kernels, we merge a set of
kernels into a megakernel---a single, monolithic kernel.  The
megakernel is launched with as many workgroups as can be occupant
concurrently.  One workgroup takes the role of the
scheduler,\footnote{\TSAdded{We note that the scheduler requirements
    given in \mysec{\ref{sec:cooperativekernels}} are agnostic to
    whether the scheduling logic takes place on the CPU or GPU. To
    avoid expensive communication between GPU and host, we choose to
    implement the scheduler on the GPU.}  } and the scheduling logic
is embedded as part of the megakernel.  The remaining workgroups act
as a pool of workers.  A worker repeatedly queries the scheduler to be
assigned a task.  A task corresponds to executing a cooperative or
non-cooperative kernel.  In the non-cooperative case, the workgroup
executes the relevant kernel function uninterrupted, then awaits
further work.  In the cooperative case, the workgroup either starts
from the kernel entry point or immediately jumps to a designated point
within the kernel, depending on whether the workgroup is an initial
workgroup of the kernel, or a forked workgroup.  In the latter case,
the new workgroup also receives a struct containing the values of all
relevant $\transmit$-annotated variables.

\myparagraph{Simplifying Assumptions}
For ease of implementation, our prototype supports multitasking a
single cooperative kernel with a single non-cooperative kernel (though
the non-cooperative kernel can be invoked many times).
We require that $\offerkill$, $\offerfork$ and
$\resizingglobalbarrier$ are called from the entry function of a
cooperative kernel.  This allows us to use $\keyword{goto}$ and
$\keyword{return}$ to direct threads into and out of the kernel.  With
these restrictions we can experiment with interesting irregular
algorithms (see \mysec\ref{sec:experiments}).  A non-mock
implementation of cooperative kernels would not use the megakernel
approach, so we did not deem the engineering effort associated with
lifting these restrictions in our prototype to be worthwhile.

%

\subsection{Scheduler Design}\label{sec:resizingbarrier}\label{sec:schedulerimpl}

To enable multitasking through cooperative kernels, the runtime (in
our case, the megakernel) must track the state of workgroups,
i.e.\ whether a workgroup is waiting or computing a kernel; maintain
consistent context states for each kernel, e.g.\ tracking the number
of active workgroups; and provide a safe way for these states to be
modified in response to $\offerfork$/$\offerkill$. We discuss these
issues, and describe the implementation of an efficient resizing
barrier. We describe how the scheduler would handle arbitrary
combinations of kernels, though as noted above our current
implementation is restricted to the case of two kernels.

\myparagraph{Scheduler Contexts}
%
%
To dynamically manage workgroups executing cooperative kernels, our
framework must track the state of each workgroup and provide a channel
of communication from the scheduler workgroup to workgroups executing
$\offerfork$ and $\offerkill$. To achieve this, we use a
\emph{scheduler context} structure, mapping a primitive workgroup id
the workgroup's status, which is either \emph{available} or the id of
the kernel that the workgroup is currently executing.  The scheduler
can then send cooperative kernels a \emph{resource message},
commanding workgroups to exit at $\offerkill$, or spawn additional
workgroups at $\offerfork$. Thus, the scheduler context needs a
communication channel for each cooperative kernel. We implement the
communication channels using atomic variables in global memory.

\myparagraph{Launching Kernels and Managing Workgroups}
To launch a kernel, the host sends a data packet to the GPU scheduler
consisting of a kernel to execute, kernel inputs, and a flag
indicating whether the kernel is cooperative. In our prototype,
this host-device communication channel is built using fine-grained SVM
atomics.

On receiving a data packet describing a kernel launch $K$, the GPU
scheduler must decide how to schedule $K$. Suppose $K$ requests $N$
workgroups. The scheduler queries the scheduler context.  If there are
at least $N$ available workgroups, $K$ can be scheduled
immediately. Suppose instead that there are only $N_a < N$ available
workgroups, but a cooperative kernel $K_c$ is executing. The scheduler
can use $K_c$'s channel in the scheduler context to command $K_c$ to
provide $N - N_a$ workgroups via $\offerkill$.  Once $N$ workgroups
are available,
the scheduler then sends $N$ workgroups from the available workgroups
to execute kernel $K$.
If the new kernel $K$ is itself a cooperative kernel, the scheduler
would be free to provide $K$ with fewer than $N$ active workgroups
initially.

If a cooperative kernel $K_c$ is executing with fewer workgroups than
it initially requested, the scheduler may decide make extra workgroups
available to $K_c$, to be obtained next time $K_c$ calls $\offerfork$.
To do this, the scheduler asynchronously signals $K_c$ through $K_c$'s
channel to indicate the number of workgroups that should join at the
next $\offerfork$ command.  When a workgroup $w$ of $K_c$ subsequently
executes $\offerfork$, thread 0 of $w$ updates the kernel and
scheduler contexts so that the given number of new workgroups are
directed to the program point after the $\offerfork$ call.  This
involves selecting workgroups whose status is \emph{available}, as
well as copying the values of $\transmit$-annotated variables to the
new workgroups.

\myparagraph{An Efficient Resizing Barrier}
In \mysec\ref{sec:semantics}, we defined the semantics of a resizing
barrier in terms of calls to other primitives.  It is possible,
however, to implement the resizing barrier with only one call to a
global barrier with $\offerfork$ and $\offerkill$ inside.

We consider barriers that use the master/slave model~\cite{XF10}: one
workgroup (master) collects signals from the other workgroups (slaves)
indicating that they have arrived at the barrier and are waiting for a
reply indicating that they may leave the barrier. Once the master has
received a signal from all slaves, it replies with a signal saying that
they may leave.

Incorporating $\offerfork$ and $\offerkill$ into such a barrier is
straightforward. Upon entering the barrier, the slaves first execute
$\offerkill$, possibly exiting. The master then waits for $M$ slaves
(the number of active workgroups), which may decrease due to
$\offerkill$ calls by the slaves, but will not increase. Once the
master observes that $M$ slaves have arrived, it knows that all other
workgroups are waiting to be released. The master executes
$\offerfork$, and the statement immediately following this
$\offerfork$ is a conditional that forces newly spawned workgroups to
join the slaves in waiting to be released. Finally, the master
releases all the slaves: the original slaves and the new slaves that
joined at $\offerfork$.

This barrier implementation is sub-optimal because workgroups only
execute $\offerkill$ once per barrier call and, depending on order of
arrival, it is possible that only one workgroup is killed per barrier
call, preventing the scheduler from gathering workgroups quickly.

We can reduce the gather time by providing a new
$\keyword{query}$ function for cooperative kernels, which returns the
number of workgroups that the scheduler needs to obtain from the
cooperative kernel.
%
%
A resizing barrier can now be implemented as follows: (1) the master
waits for all slaves to arrive; (2) the master calls $\offerfork$ and
commands the new workgroups to be slaves; (3) the master calls
$\keyword{query}$, obtaining a value $W$; (4) the master releases the
slaves, broadcasting the value $W$ to them; (5) workgroups with ids
larger than $M-W$ spin, calling $\offerkill$ repeatedly until the
scheduler claims them---we know from $\keyword{query}$ that the
scheduler will eventually do so.
We show in
\mysec\ref{sec:responsiveness} that the barrier using $\keyword{query}$ greatly
reduces the gather time in practice.

\section{Applications and Experiments}\label{sec:experiments}

We discuss our experience porting irregular algorithms to cooperative
kernels and describe the GPUs on which we evaluate these applications
(\mysec\ref{sec:portingalgorithms}).  For these GPUs, we report on
experiments to determine non-cooperative workloads that model the
requirements of various graphics rendering tasks
(\mysec\ref{sec:sizingnoncoop}).  We then examine the overhead
associated with moving to cooperative kernels when multitasking is
\emph{not} required (\mysec\ref{sec:overhead}), as well as the
responsiveness and throughput observed when a cooperative kernel is
multi-tasked with non-cooperative workloads
(\mysec\ref{sec:responsiveness}). Finally, we compare
against a performance model of \emph{kernel-level} preemption, which
we understand to be what current Nvidia GPUs provide (\mysec\ref{sec:nvidiacomparison}).


\subsection{Applications and GPUs}\label{sec:portingalgorithms}

\begin{table}[t]
\normalsize
\caption{Blocking GPU applications investigated}
\centering
\begin{tabular}{ l r r r r r r}
App. & barriers & kill & fork & transmit & LoC & inputs\\
\hline
\rowcolor{Gray1}
color & 2 / 2 & 0 & 0 & 4 & 55 & 2\\
\rowcolor{Gray1}
mis & 3 / 3 & 0 & 0 & 0 & 71 & 2\\
\rowcolor{Gray1}
p-sssp & 3 / 3 & 0 & 0 & 0  & 42 & 1\\
\rowcolor{Gray2}
bfs & 2 / 2 & 0 & 0  & 4  & 185 & 2\\
\rowcolor{Gray2}
l-sssp & 2 / 2 & 0 & 0  & 4  & 196 & 2\\
\rowcolor{Gray3}
octree & 0 / 0 & 1 & 1 & 0 & 213 & 1 \\
\rowcolor{Gray3}
game & 0 / 0 & 1 & 1 & 0 & 308 & 1 \\
\end{tabular} \\
\crule[Gray1]{.2cm}{.2cm} Pannotia \hspace{.4cm} \crule[Gray2]{.2cm}{.2cm} Lonestar GPU  \hspace{.4cm}  \crule[Gray3]{.2cm}{.2cm} work stealing

\label{tab:applications}
\end{table}


\mytablong\ref{tab:applications} gives an overview of the 7 irregular
algorithms that we ported to cooperative kernels. Among them, 5 are
graph algorithms, based on the Pannotia~\cite{Pannotia} and
Lonestar~\cite{BNP12} GPU application suites, using global barriers.
We indicate how many of the original number of barriers are changed to
resizing barriers (all of them), and how many variables need to be
transmitted.  The remaining two algorithms are work stealing
applications: each required the addition of $\offerfork$ and
$\offerkill$ at the start of the main loop, and no variables needed to
be transmitted (similar to example discussed in \mysec\ref{sec:programmingguidelines}).
Most graph applications come with 2 different data sets as input,
leading to 11 application/input pairs in total.


\TSAdded{Our prototype implementation (\mysec\ref{sec:implementation}) requires
two optional features of OpenCL 2.0: SVM fine-grained buffers and SVM
atomics. Out of the GPUs available to us, from ARM, AMD, \nvidia, and
Intel, only Intel GPUs provided robust support of these features.}


%
%
\TSAdded{We thus ran our experiments on three Intel GPUs: HD 520, HD 5500
and Iris 6100. The results were similar across the GPUs, so for
conciseness, we report only on the Iris 6100 GPU (driver
20.19.15.4463) with a host CPU i3-5157U. The Iris has a reported 47
compute units. Results for the other Intel GPUs are presented in Appendix~\ref{appendix:extragraphs}.}

\subsection{Sizing Non-cooperative Kernels}\label{sec:sizingnoncoop}

Enabling rendering of smooth graphics in parallel with irregular
algorithms is an important use case for our approach.  Because our
prototype implementation is based on a megakernel that takes over the
entire GPU (see \mysec\ref{sec:implementation}), we cannot assess this
directly.

We devised the following method to determine OpenCL workloads that simulate
the computational intensity of various graphics rendering workloads.
We designed a synthetic kernel that occupies all workgroups of a GPU
for a parameterised time period $t$, invoked in an infinite loop by a
host application.  We then searched for a maximum value for $t$ that
allowed the synthetic kernel to execute without having an observable
impact on graphics rendering.  Using the computed value, we ran the
host application for $X$ seconds, measuring the time $Y < X$ dedicated
to GPU execution during this period and the number of kernel launches
$n$ that were issued.  We used $X \geq 10$ in all experiments.  The
values $(X-Y)/n$ and $X/n$ estimate the average time spent using the
GPU to render the display between kernel calls (call this $E$) and the
period at which the OS requires the GPU for display rendering (call
this $P$), respectively.

We used this approach to measure the GPU availability required for three
types of rendering: \emph{light}, whereby desktop icons are smoothly
emphasised under the mouse pointer; \emph{medium}, whereby window
dragging over the desktop is smoothly animated; and \emph{heavy}, which
requires smooth animation of a WebGL shader in a browser.  For
\emph{heavy} we used WebGL demos from the Chrome
experiments~\cite{chrome-experiments}.



Our results are the following: $P=70\mathit{ms}$ and $E=3\mathit{ms}$ for light;
$P=40\mathit{ms}$, $E=3\mathit{ms}$ for medium; and $P=40\mathit{ms}$, $E=10\mathit{ms}$ for heavy. For
medium and heavy, the $40\mathit{ms}$ period coincides with the human persistence
of vision. The $3\mathit{ms}$ execution duration of both light and medium
configurations indicates that GPU computation is cheaper for basic
display rendering compared with more complex rendering.




\subsection{The Overhead of Cooperative Kernels}\label{sec:overhead}

\myparagraph{Experimental Setup}
Invoking the cooperative scheduling primitives incurs some overhead
even if no killing, forking or resizing actually occurs, because the cooperative kernel still needs to interact with the scheduler to determine this.
We assess this overhead by measuring the
 slowdown in execution time between the original and cooperative versions of a kernel, forcing the scheduler to never modify the number of
active workgroups in the cooperative case.

Recall that our mega kernel-based implementation merges the code of a
cooperative and a non-cooperative kernel.
This can reduce the occupancy for the merged kernel, e.g.\ due to
higher register pressure, This is an artifact of our prototype
implementation, and would not be a problem if our approach was
implemented inside the GPU driver.  We thus launch both the original
and cooperative versions of a kernel with the reduced occupancy bound
in order to meaningfully compare execution times.

\begin{figure*}
\includegraphics[width=.67\columnwidth]{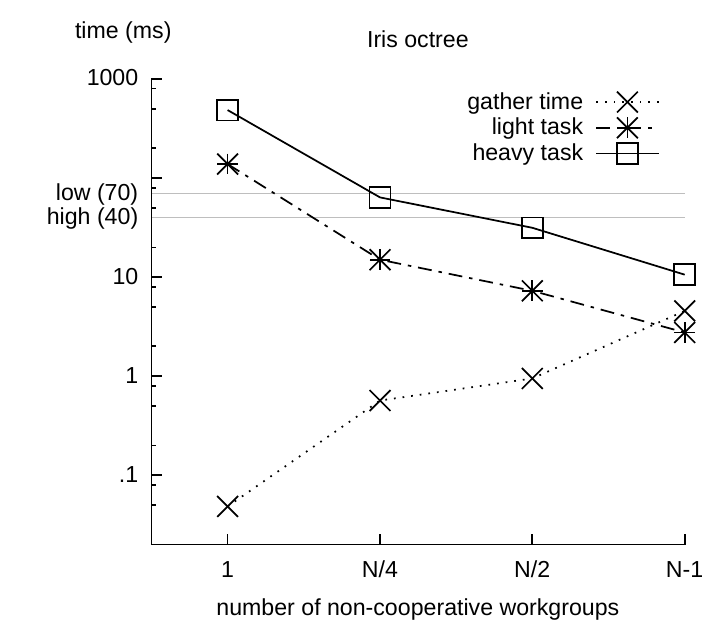}
\includegraphics[width=.67\columnwidth]{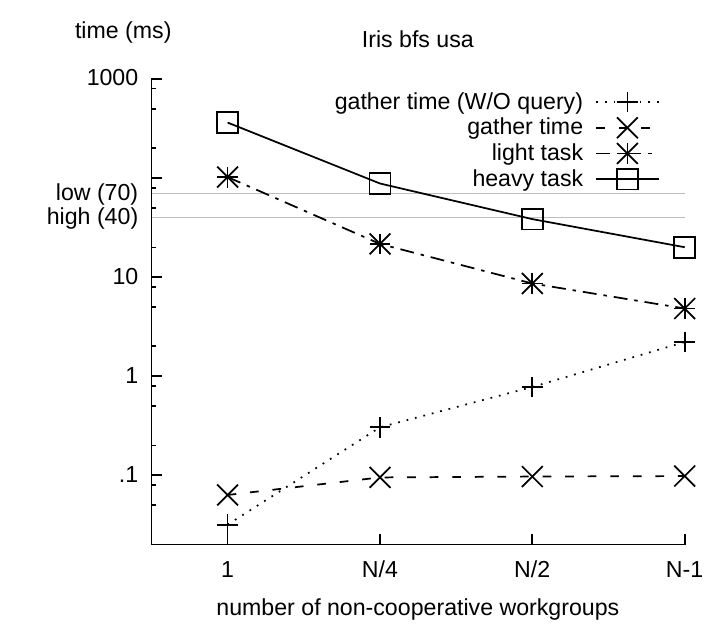}
\includegraphics[width=.67\columnwidth]{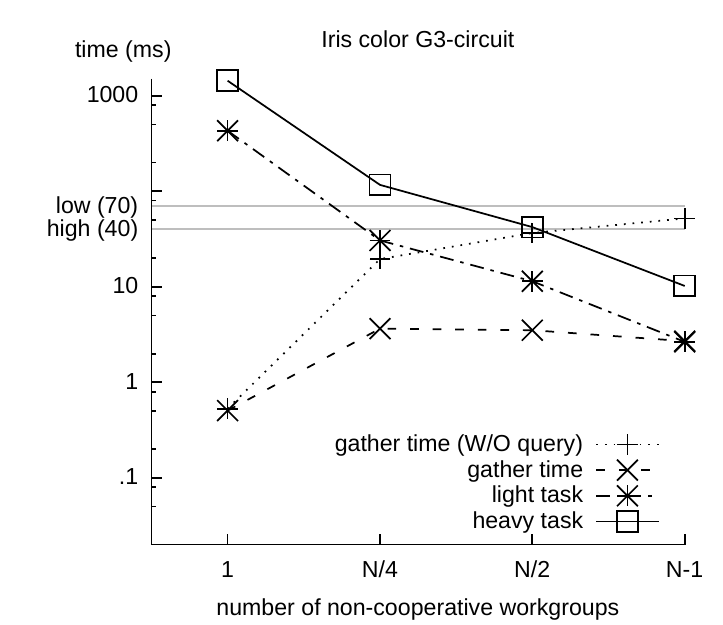}
\caption{Example gather time and non-cooperative timing results}\label{fig:fine-grained-timing}
\end{figure*}

\begin{table}
\normalsize
\caption{Cooperative kernel slowdown w/o multitasking}
\centering
\begin{tabular}{ l l | l l | l l }
\multicolumn{2}{c|}{overall} & \multicolumn{2}{c|}{barrier} & \multicolumn{2}{c}{wk.steal.} \\
mean & max & mean & max & mean & max \\
\hline
$1.07$ & $1.23^{\ddagger}$ & $1.06$ & $1.20^{\diamond}$ & $1.12$ & $1.23^{\ddagger}$ \\
\end{tabular}\\
{\small
 $^{\ddagger}$octree, $^{\diamond}$color G3\_circuit
}
\label{tab:overhead}
\end{table}

\myparagraph{Results}
\mytab\ref{tab:overhead} shows the geometric mean and
maximum slowdown across all applications and inputs, with averages and
maxima computed over 10 runs per benchmark. For the maximum slowdowns,
we indicate which application and input was responsible. The slowdown is
below 1.25 even in the worst case, and closer to 1 on average. We consider
these results encouraging, especially since the performance of our
prototype could clearly be improved upon in a native implementation.

\subsection{Multitasking via Cooperative Scheduling}\label{sec:responsiveness}

We now assess the responsiveness of multitasking between a
long-running cooperative kernel and a series of short, non-cooperative
kernel launches, and the performance impact of multitasking on the
cooperative kernel.


\myparagraph{Experimental Setup} For a given cooperative kernel and
its input, we launch the kernel and then repeatedly schedule a
non-cooperative kernel that aims to simulate the intensity of one of
the three classes of graphics rendering workload discussed in
\mysec\ref{sec:sizingnoncoop}. In practice, we use matrix
multiplication as the non-cooperative workload, with matrix
dimensions tailored to reach the appropriate execution duration.  We
conduct separate runs where we vary the number of workgroups requested
by the non-cooperative kernel, considering the cases where one, a
quarter, a half, and all-but-one, of the total number of workgroups
are requested. For the graph algorithms we try both
regular and query barrier implementations.

Our experiments span 11 pairs of cooperative kernels and inputs, 3
classes of non-cooperative kernel workloads, 4 quantities of
workgroups claimed for the non-cooperative kernel and 2 variations of
resizing barriers for graph algorithms, leading to 240 configurations.
We run each configuration 10 times, in order to report averaged
performance numbers. For each run, we record the execution time of the
cooperative kernel. For each scheduling of the non-cooperative kernel
during the run, we also record the \emph{gather time} needed by the
scheduler to collect workgroups to launch the non-cooperative kernel,
and the non-cooperative kernel execution time.
%



\myparagraph{Responsiveness}
\myfiglong\ref{fig:fine-grained-timing} reports, on three
configurations, the average gather and execution times for the
non-cooperative kernel with respect to the quantity of workgroups allocated to
it.  A logarithmic scale is used for time since gather times tend to
be much smaller than execution times. The horizontal grey lines
indicates the desired period for non-cooperative kernels.  These
graphs show a representative sample of our results; the full set of
graphs for all configurations is provided in Appendix~\ref{appendix:extragraphs}.

The left-most graph illustrates a work
stealing example.  When the non-cooperative kernel is given only one
workgroup, its execution is so long that it cannot complete within the
period required for a screen refresh. The gather time is very good
though, since the scheduler needs to collect only one workgroup.  The
more workgroups are allocated to the non-cooperative kernels, the
faster it can compute: here the non-cooperative kernel becomes fast
enough with a quarter (resp.\ half) of available workgroups for light
(resp.\ heavy) graphics workload. Inversely, the gather time increases
since the scheduler must collect more and more workgroups.

The middle and right graphs show results for graph algorithms.  These
algorithms use barriers, and we experimented with the regular and
query barrier implementations described in
\mysec\ref{sec:resizingbarrier}.  The execution times for the
non-cooperative task are averaged across all runs, including with both
types of barrier.  We show separately the average gather time
associated with each type of barrier.  The graphs show a similar trend
to the left-most graph: as the number of non-cooperative workgroups
grows, the execution time decreases and the gather time
increases. 
%
The gather time is higher on the rightmost figure as the G3 circuit input
graph is rather wide than deep, so the graph algorithm reaches
resizing barriers less often than for the USA road input of the middle
figure for instance. The scheduler thus has fewer opportunities to
collect workgroups and gather time increases. Nonetheless, scheduling
responsiveness can benefit from the query barrier: when used, this
barrier lets the scheduler collect all needed workgroups as soon as
they hit a resizing barrier.
As we can see, the gather time of the
query barrier is almost stable with respect to the number of workgroups that
needs to be collected.

\begin{figure}
\includegraphics[width=\columnwidth]{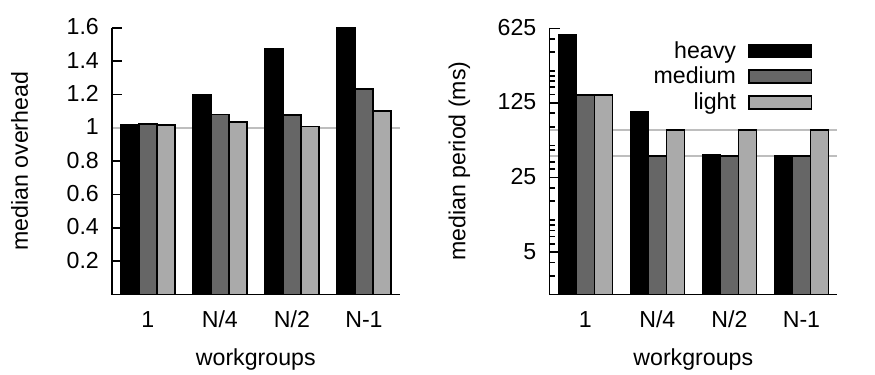}
\caption{Performance impact of multitasking cooperative and non-cooperative workloads, and the period with which non-cooperative kernels execute}\label{fig:performance}
\end{figure}

\myparagraph{Performance} \myfiglong\ref{fig:performance} reports the
overhead brought by the scheduling of non-cooperative kernels over the
cooperative kernel execution time.  This is the slowdown associated
with running the cooperative kernel in the presence of multitasking,
vs.\ running the cooperative kernel in isolation (median over all
applications and inputs).  We also show the period at which
non-cooperative kernels can be scheduled (median over all applications
and inputs). Our data included some outliers that occur with
benchmarks in which the resizing barrier are not called very
frequently and the graphics task requires half or more workgroups. For
example, a medium graphics workload for bfs on the rmat input has over
an 8$\times$ overhead when asking for all but one of the
workgroups. As \myfiglong\ref{fig:performance} shows, most of our
benchmarks are much better behaved than this. In future work
is required to examine the problematic benchmarks in more detail,
possibly inserting more resizing calls.

We show results for the three workloads listed in
\mysec~\ref{sec:sizingnoncoop}. The horizontal lines in the period
graph correspond to the goals of the workloads: the higher
(resp. lower) line corresponds to a period of $70\mathit{ms}$ (resp.\ $40\mathit{ms}$) for
the light (resp. medium and heavy) workload.

Co-scheduling non-cooperative kernels that request a single workgroup
leads to almost no overhead, but the period is far too high to meet
the needs of any of our three workloads; e.g.\ a heavy workload
averages a period of $939\mathit{ms}$. As more workgroups are dedicated to
non-cooperative kernels, they execute quickly enough to be scheduled
at the expected period. For the light and medium workloads, a quarter
of the workgroups executing the non-cooperative kernel are able to
meet their goal period (70 and $40\mathit{ms}$ resp.). However, this is not
sufficient to meet the goal for the heavy workload (giving a median
period of $104\mathit{ms}$). If half of the workgroups are allocated to the
non-cooperative kernel, the heavy workload achieves its goal period
(median of $40\mathit{ms}$).
Yet, as expected, allocating more non-cooperative workgroups increases
the overhead of the cooperative kernel.
Still, heavy workloads meet their period by allocating half
of the workgroups, incurring a slow down of less than
1.5$\times$ (median). Light and medium workloads meet their period
with only a small overhead; 1.04$\times$ and 1.08$\times$ median
slowdown respectively.


\subsection{Comparison with Kernel-Level Preemption}\label{sec:nvidiacomparison}

\begin{table}[t]
\normalsize
\caption{Overhead of kernel level preemption vs cooperative kernels for three
graphics workloads}

\centering
\begin{tabular}{ l r r r}
g. workload &  kernel-level  & cooperative & resources \\
\hline
light & 1.04 & 1.04 & $N/4$\\
medium & 1.08 & 1.08 & $N/4$\\
heavy & 1.33 & 1.47 & $N/2$\\
\end{tabular} \\

\label{tab:preemption}
\end{table}

Nvidia's recent Pascal architecture provides hardware support for
instruction-level preemption~\cite{PascalWhitepaper,anandtech},
however, preemption of entire kernels, but not of individual
workgroups is supported.  Intel GPUs do not provide this feature, and
our OpenCL prototype of cooperative kernels cannot run on Nvidia GPUs,
making a direct comparison impossible.  We present here a theoretical
analysis of the overheads associated with sharing the GPU between
graphics and compute tasks via kernel-level preemption.

Suppose a graphics workload is required to be scheduled with period
$P$ and duration $D$, and that a compute kernel requires time $C$ to
execute without interruption.  If we assume the cost of preemption is
negligible (e.g.\ Nvidia have reported preemption times of 0.1
$\mathit{ms}$ for Pascal~\cite{anandtech}, because of
special hardware support), then the overhead associated with switching
between compute and graphics every $P$ time steps is $P/(P-D)$.

We compare this task-level preemption overhead model with our
experimental results per graphics workload in
\mytab{\ref{tab:preemption}}. We report the overhead of the
configuration that allowed us to meet the deadline of the graphics
task.
Based on the above assumptions, our approach provides similar overhead
for low and medium graphics workloads, however, has a higher overhead for
the high workload.

Our low performance for heavy workloads is because the graphics task
requires half of the workgroups, crippling the cooperative kernel
enough that $\offerfork$ calls are not issued as frequently.  Future
work may examine how to insert more resizing calls in these
applications to address this.
These results suggest that a hybrid preemption scheme may work
well. That is, the cooperative approach works well for light and
medium tasks; on the other hand, heavy graphics tasks benefit from the
coarser grained, kernel-level preemption strategy. However, the
preemption strategy requires specialised hardware support
in order to be efficient.

\section{Related Work}\label{sec:relatedwork}

\myparagraph{Irregular Algorithms and Persistent kernels}
There has been a lot of work on accelerating blocking irregular
algorithms using GPUs, and on the \emph{persistent threads}
programming style for long-running
kernels~\cite{owens-persistent,DBLP:conf/ipps/KaleemVPHP16,DBLP:conf/ipps/DavidsonBGO14,DBLP:conf/hipc/HarishN07,DBLP:journals/topc/MerrillGG15,DBLP:conf/egh/VineetHPN09,DBLP:conf/ppopp/NobariCKB12,DBLP:conf/hpcc/SolomonTT10a,DBLP:conf/popl/PrabhuRMH11,DBLP:conf/ppopp/Mendez-LojoBP12,DBLP:conf/oopsla/PaiP16,DBLP:conf/oopsla/SorensenDBGR16,DBLP:conf/egh/CedermanT08,TPO10,BNP12,Pannotia}.
These approaches rely on the occupancy-bound execution model, flooding
available compute units with work, so that the GPU is unavailable for
other tasks, and assuming fair scheduling between occupant workgroups,
which is unlikely to be guaranteed on future GPU platforms.
As our experiments demonstrate, our cooperative kernels model allows blocking algorithms
to be upgraded to run in a manner that facilitates responsive multitasking.


\myparagraph{GPU Multitasking and Scheduling}
%
%
Hardware support for preemption has been proposed for \nvidia GPUs, as
well as \emph{SM-draining} whereby workgroups occupying a symmetric
multiprocessor (SM; a compute unit using our terminology) are allowed
to complete until the SM becomes free for other
tasks~\cite{DBLP:conf/isca/TanasicGCRNV14}.  SM draining is limited
the presence of blocking constructs, since it may not be possible to
drain a blocked workgroup.
A follow-up work adds the notion of SM \emph{flushing}, where a
workgroup can be re-scheduled from scratch if it has not yet committed
side-effects~\cite{DBLP:conf/asplos/ParkPM15}.  Both approaches have
been evaluated using simulators, over sets of regular GPU kernels.
Very recent \nvidia GPUs (i.e. the Pascal architecture) support
preemption, though, as discussed in \mysec{\ref{sec:intro}} and
\mysec{\ref{sec:nvidiacomparison}}, it is not clear whether they guarantee
fairness or allow tasks to share GPU resources at the workgroup
level~\cite{PascalWhitepaper}.

CUDA and OpenCL provide the facility for a kernel to spawn further
kernels~\cite{cuda-75}.  This \emph{dynamic parallelism} can be used
to implement a GPU-based scheduler, by having an initial scheduler
kernel repeatedly spawn further kernels as required, according to some
scheduling policy~\cite{DBLP:conf/ppopp/Muyan-OzcelikO16}.  However,
kernels that uses dynamic parallelism are still prone to unfair
scheduling of workgroups, and thus does not help in deploying blocking
algorithms on GPUs.

\myparagraph{Cooperative Multitasking}
Cooperative multitasking was offered in older operating systems
(e.g. pre 1995 Windows) and is still used by some operating systems,
such as RISC OS~\cite{risc-os-multitasking}. \TSAdded{Additionally,
  cooperative multitasking can be efficiently implemented in today's
  high-level languages for domains in which preemptive multitasking is
  either too costly or not supported on legacy
  systems~\cite{Tarpenning:1991:CMC:136810.136820}}.


%


\section{Conclusions and Future Work}\label{sec:conclusion}

We have proposed \emph{cooperative kernels}, a small set of GPU
programming extensions that allow long-running, blocking kernels to be
fairly scheduled and to share GPU resources with other workloads.
Experimental results using our megakernel-based prototype show that
the model is a good fit for current GPU-accelerated irregular
algorithms.  The performance that could be gained through a native
implementation with driver support would be even better.
Avenues for future work include seeking additional classes of
irregular algorithms to which the model might (be extended to) apply
(to), investigating implementing native support in open source
drivers, and integrating cooperative kernels into template- and
compiler-based programming models for graph algorithms on
GPUs~\cite{DBLP:conf/ppopp/WangDPWRO16,DBLP:conf/oopsla/PaiP16}.

\section*{Acknowledgments}
We are grateful to Lee Howes, Bernhard Kainz, Paul Kelly, Christopher
Lidbury, Steven McDonagh, Sreepathi Pai, and Andrew Richards for
insightful comments throughout the work. We thank the FSE reviewers
for their thorough evaluations and feedback. This work is supported in
part by EPSRC Fellowship EP/N026314, and a gift from Intel Corporation.


\clearpage

\appendix

\section{Operational Semantics for Cooperative Kernels}\label{appendix:semantics}

\newcommand{\myss}{\mathit{ss}}
\newcommand{\Stmts}{\mathsf{Stmts}}
\newcommand{\threadstates}{\mathsf{ThreadStates}}
\newcommand{\sharedstates}{\mathsf{SharedStates}}
\newcommand{\sync}{\mathsf{sync}}

In \mysec\ref{sec:semantics} we presented the semantics of cooperative
kernels relatively informally, using English, to provide the intuition
behind our programming model.  We now back this up with a more formal
presentation as an operational semantics for an abstract GPU
programming model.

\myparagraph{States}
Let $L$ be a set of \emph{local states} that abstractly captures the
private memory associated with a thread executing a GPU kernel.  Let
$\Stmts$ denote the set of all possible statements that a
thread can execute.  We do not detail the structure of these
statements, except that we assume sequential composition of statements
is provided by the $\code{;}$ separator, and that the $\offerkill$,
$\offerfork$, $\globalbarrier$ and $\resizingglobalbarrier$ primitives
from our cooperative kernels programming model are valid statements.

A \emph{thread state} is then a pair $(l, \myss)$, where $l \in L$ and
$\myss \in \Stmts$.  The $l$ component captures the valuation of all the
thread's private memory, and the $\myss$ component captures the
remaining statements to be executed by the thread.  Let $\threadstates$ denote the set of all thread states.

Assuming that $d > 0$ threads per workgroup were requested on kernel launch, a \emph{workgroup state}
is a $d$-tuple $((l_0, \myss_0), \dots, (l_{d-1}, \myss_{d-1}))$, where each $(l_i, \myss_i)$ is the thread state for the $i$th thread in the workgroup $(0\leq i < d$).

Assuming that $N > 0$ was specified as the maximum number of
workgroups that should execute the cooperative kernel, a \emph{kernel
  state} is then a pair
\[
(\sigma, (w_0, \dots, w_{M-1}, \underbrace{\bot, \dots,
\bot}_{N-M}))\]
where: $\sigma$ represents the shared state of the kernel; $M \leq N$
is the number of \emph{active} workgroups; $w_i$ is the workgroup
state for active workgroup $i$ ($0 \leq i < M$); and $N-M$ occurrences
of $\bot$ indicate \emph{absent} workgroups.  Let $\sharedstates$
denote the set of all possible shared states.  We regard
workgroup-local storage as being part of the shared state of a kernel.

\myparagraph{Thread-level transitions}
We leave the semantics for thread-level transitions abstract, assuming
a binary relation $\rightarrow_{tau}$ on $\sharedstates{} \times
\threadstates{}$.  If $(\sigma, (l, \myss)) \rightarrow_{\tau}
(\sigma', (l', \myss'))$, this indicates that if a thread is in local
state $(l, \myss)$, the thread can transition to local state $(l',
\myss')$, changing the shared state from $\sigma$ to $\sigma'$ in the
process.

All we require is that $(\sigma, (l, \myss)) \rightarrow_{\tau}
(\sigma', (l', \myss'))$ if $\myss$ has the form $\mathsf{special}();
\mathit{tt}$, where $\mathsf{special}$ is one of $\offerkill$,
$\offerfork$, $\globalbarrier$ or $\resizingglobalbarrier$.  This is
because we shall specifically define the meaning of the new primitives
introduced by our programming model.

\myparagraph{Memory synchronisation}
GPUs are known to have relaxed memory models~\cite{ABDGKPSW-2015}.  To abstractly
account for this, we assume that the shared state component $\sigma$
is not simply a mapping from locations to values, but instead
captures all the intricacies of GPU memory that can lead to this
relaxed behaviour.  We also assume a function $\sync$ which, given a
kernel state $\kappa$, returns a set of kernel states.  The idea is
that each $\kappa' \in \sync(\kappa)$ is a possible kernel state that
can be reached from $\kappa$ by stalling until all stores to memory
and loads from memory to thread-local storage have completed.  All we
require is that $\sync$ does not modify the number of active
workgroups nor the component of a thread state that determines which
statements remain to be executed.

\begin{figure*}
\center
\[
$$
\inferrule{
w_i(j) = (l, \myss)
\\
(\sigma, (l, \myss)) \rightarrow_{\tau} (\sigma', (l', \myss'))
\\
w_i' = w_i[j \mapsto (l', \myss')]
}
{
(\sigma, (\dots, w_i, \dots)) \rightarrow (\sigma', (\dots, w_i', \dots))
}
(\textsc{Thread-Step})
$$
\]

\medskip

\[
$$
\inferrule{
\forall j \;.\; w_i(j) = (l_j, \offerkill();\myss)
\\
w_i' = ((l_0, \myss), \dots, (l_{d-1}, \myss))
}
{
(\sigma, (\dots, w_i, \dots)) \rightarrow (\sigma, (\dots, w_i', \dots))
}
(\textsc{Kill-No-Op})
$$
\]

\medskip

\[
$$
\inferrule{
\forall j \;.\; w_{M-1}(j) = (l_j, \offerkill();\myss)
\\
M > 0
}
{
(\sigma, (\dots, w_{M-2}, w_{M-1}, \bot, \dots, \bot)) \rightarrow (\sigma, (\dots, w_{M-2}, \bot, \bot, \dots, \bot))
}
(\textsc{Kill})
$$
\]

\medskip

\[
$$
\inferrule{
\forall j \;.\; w_i(j) = (l_j, \offerfork();\myss)
\\
w_i' = ((l_{0}, \myss), \dots, (l_{d-1}, \myss))
\\
k \in [0, N - M]
\\
\forall a \in [0, k - 1] \;.\; w_{M+a} = ((l_{0}, \myss), \dots, (l_{0}, \myss))
}
{
(\sigma, (\dots, w_i, \dots, w_{M-1}, \bot, \dots, \bot)) \rightarrow (\sigma, (\dots, w_i', \dots, w_{M-1}, w_{M}, \dots, w_{M+k-1}, \bot, \dots, \bot))
}
(\textsc{Fork})
$$
\]

\medskip

\[
$$
\inferrule{
\forall i \;.\;\forall j\;.\;w_i(j) = (l_{i,j}, \globalbarrier();\myss)
\\
\forall i \;.\;\forall j\;.\;w_i'(j) = (l_{i,j}, \myss)
\\\\
\kappa \in \sync((\sigma , (w_{0}', \dots, w_{M-1}', \bot, \dots, \bot))
}
{
(\sigma, (w_{0}, \dots, w_{M-1}, \bot, \dots, \bot)) \rightarrow \kappa
}
(\textsc{Barrier})
$$
\]

\medskip

\[
$$
\inferrule{
\forall i \;.\;\forall j\;.\;w_i(j) = (l_{i,j}, \resizingglobalbarrier();\myss)
\\
\forall j\;.\;w_{0}'(j) = (l_{1,j}, \globalbarrier(); \offerfork(); \globalbarrier(); \globalbarrier();\myss)
\\
\forall i \neq 1 \;.\;\forall j\;.\;w_i'(j) = (l_{i,j}, \globalbarrier(); \globalbarrier(); \offerkill(); \globalbarrier();\myss)
}
{
(\sigma, (w_{0}, \dots, w_{M-1}, \bot, \dots, \bot)) \rightarrow (\sigma, (w_{0}', \dots, w_{M-1}', \bot, \dots, \bot))
}
(\textsc{Resizing-Barrier})
$$
\]

\caption{Abstract operational semantics for our cooperative kernels language extensions}\label{fig:semanticrules}

\end{figure*}

\myparagraph{Operational semantics rules}
\myfiglong\ref{fig:semanticrules} presents the rules of our
operational semantics, defining a relation $\rightarrow$ on kernel
states.

Rule $\textsc{Thread-Step}$ defines the semantics for thread making a
single execution step, delegating to the abstract $\rightarrow_{\tau}$
relation to determine how the thread's local state and the shared
state component change.  For simplicity, this rule ignores the
semantics of intra-workgroup barriers, which are not our focus here.

Rule $\textsc{Kill-No-Op}$ reflects the fact that when a workgroup
reaches $\offerkill$, this can always be treated as a no-op.  Whether
a scheduler implementation accepts $\offerkill$ calls or not depends
on competing workloads and how the scheduler has been designed to meet
the non-functional requirements discussed in
\mysec\ref{sec:nonfunctional}, but in general the programmer should
always be prepared for the possibility that a workgroup survives after
calling $\offerkill$.

The case where a workgroup's offer to be killed is accepted by the
scheduler is captured by rule $\textsc{Kill}$.  Because we have
adopted a semantics where workgroup 0 is never killed and where only
the workgroup with the highest id can be killed, the rule only fires
if $M > 0$ and workgroup $M-1$ has reached $\offerkill$.  The rule has
the effect of replacing the workgroup state for $w_{M-1}$
with $\bot$.

Recall that $\offerkill$ is a workgroup-level function: the same
syntactic $\offerfork$ call must be reached by all threads in a
workgroup.  This is captured in our rules by requiring in both
$\textsc{Kill-No-Op}$ and $\textsc{Kill}$ that every thread is ready
to execute $\offerkill$ followed by an identical statement $\myss$.
Neither rule applies until all threads in a workgroup reach
$\offerfork$, and the workgroup gets stuck if multiple threads in a
workgroup reach $\offerfork$ with different subsequent statements.

The $\textsc{Fork}$ rule similarly requires all threads in a workgroup
to reach $\offerfork$ with identical following statements.  A
nondeterministic number of new workgroups, $k$, is selected to be
forked, where $k \in [0, N-M]$.  Importantly, $k=0$ is always a
legitimate choice, in which case $\offerfork$ has no effect on the
number of workgroups that are executing.  In the case where $k > 0$,
$k$ new workgroup states are created, where each workgroup inherits
the local state of thread 0 in $w_i$, the workgroup executing the fork
call.  After $\offerfork$, all threads in all workgroups, including
the new threads, proceed to execute the sequence of statements $\myss$
that followed $\offerfork$.

A simplification here is that we do not model transmission of
particular annotated variables from thread 0 of the forking workgroup,
instead specifying that the entire local state of the workgroup is
transmitted.  Extending the semantics to transmit only annotated
variables would be straightforward but verbose, requiring the local
memory component of a thread state to be split into two: the part of
the local state to be transmitted (modelling $\transmit$-annotated
variables), and the rest of the local state.

The $\textsc{Barrier}$ rule requires that \emph{every} thread
executing the kernel reaches $\globalbarrier$ an identical following
statement.  This reflects that fact that $\globalbarrier$ is a
kernel-level function.  Each thread then skips over the
$\globalbarrier$ call, and the $\sync$ function is applied to yield
a set of kernel states that can arise due to memory synchronization
taking place.  An arbitrary member of this set is selected as the next
kernel state.

Despite its apparent complexity, the $\textsc{Resizing-Barrier}$ rule simply implements the rewriting of $\resizingglobalbarrier$ in terms of $\globalbarrier$, $\offerkill$ and $\offerfork$ discussed in \mysec\ref{sec:semantics}.

\section{Alternative Semantic Choices}\label{appendix:semanticalternatives}

The semantics of cooperative kernels has been guided by the practical
applications we have studied (described in
\mysec\ref{sec:portingalgorithms}).  We now discuss several cases
where we might have taken different and also reasonable semantic
decisions.

\myparagraph{Killing order}
We opted for a semantics whereby only the active workgroup with the
highest id can be killed.  This has an appealing property: it means
that the ids of active workgroups are contiguous, which is important
for processing of contiguous data.  The cooperative graph traversal
algorithm of \myfig\ref{fig:cgraphsearch} illustrates this: the
algorithm is prepared for $\getglobalsize$ to change after each
resizing barrier call, but depends on the fact that $\getglobalid$
returns a contiguous range of thread ids.

A disadvantage of this decision is that it may provide sub-optimal
responsiveness from the point of view of the scheduler.  Suppose the
scheduler requires an additional compute unit, but the active thread
with the largest id is processing some computationally intensive work
and will take a while to reach $\offerkill$.  Our semantics means that
the scheduler cannot take advantage of the fact that another active
workgroup may invoke $\offerkill$ sooner.

Cooperative kernels that do not require contiguous thread ids might me
more suited to a semantics in which workgroups can be killed in any
order, but where workgroup ids (and thus thread global ids) are not
guaranteed to be contiguous.

\myparagraph{Keeping one workgroup alive}
Our semantics dictate that the workgroup with id 0 will not be killed
if it invokes $\offerkill$.  This avoids the possibility of the
cooperative kernel terminating early due to the programmer
inadvertently allowing all workgroups to be killed, and the decision
to keep workgroup 0 alive fits well with our choice to kill workgroups
in descending order of id.

However, there might be a use case for a cooperative kernel reaching a
point where it would be acceptable for the kernel to exit, although
desirable for some remaining computation to be performed if competing
workloads allow it.  In this case, a semantics where all workgroups can be killed via $\offerkill$ would be appropriate, and the programmer would need to guard each $\offerkill$ with an id check in cases where killing all workgroups would be unacceptable.  For example:
\lstset{basicstyle=\tt,numbers=none}
\begin{lstlisting}
  if($\getgroupid{0}$ != 0) $\offerkill$();
\end{lstlisting}
\lstset{basicstyle=\scriptsize\tt,numbers=left}
would ensure that at least workgroup 0 is kept alive.

\myparagraph{Transmission of partial state from a single thread}
Recall from the semantics of $\offerfork$ that newly forked workgroups
inherit the variable valuation associated with thread 0 of the forking
workgroup, but only for $\transmit$-annotated variables.  Alternative
choices here would be to have forked workgroups inherit values for
\emph{all} variables from the forking workgroup, and to have thread
$i$ in the forking workgroup provide the valuation for thread $i$ in
each spawned workgroup, rather than having thread 0 transmit the
valuation to all new threads.

We opted for transmitting only selected variables based on the
observation that many of a thread's private variables are dead at the
point of issuing $\offerfork$ or $\resizingglobalbarrier$, thus it
would be wasteful to transmit them.  A live variable analysis could
instead be employed to over-approximate the variables that might be
accessed by newly arriving workgroups, so that these are automatically
transmitted.

In all cases, we found that a variable that needed to be transmitted
had the property of being uniform across the workgroup.  That is,
despite each thread having its own copy of the variable, each thread
is in agreement on the variable's value.  As an example, the
$\keyword{level}$, $\keyword{in\_nodes}$ and $\keyword{out\_nodes}$
variables used in \myfig\ref{fig:cgraphsearch} are all stored in thread-private
memory, but all threads in a workgroup agree on the values of these
variables at each $\resizingglobalbarrier$ call.  As a result,
transmitting the thread 0's valuation of the annotated variables is
equivalent to (and more efficient than) transmitting values on a
thread-by-thread basis.  We have not yet encountered a real-world
example where our current semantics would not suffice.

\section{Graphs for Multitasking Experiments}\label{appendix:extragraphs}

We present the full set of graphs exemplified by the examples in
\myfig\ref{fig:fine-grained-timing}; for completeness we reproduce
those graphs here too.

\includegraphics[width=.7\columnwidth]{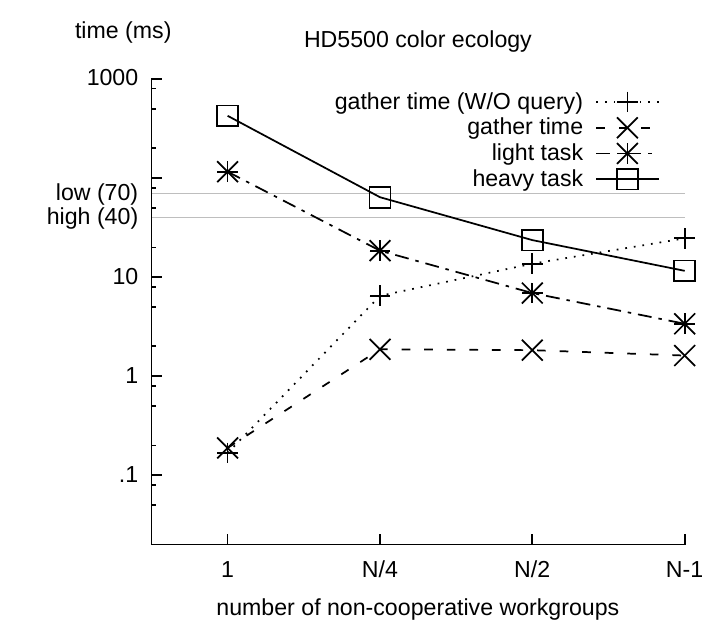} \\
\includegraphics[width=.7\columnwidth]{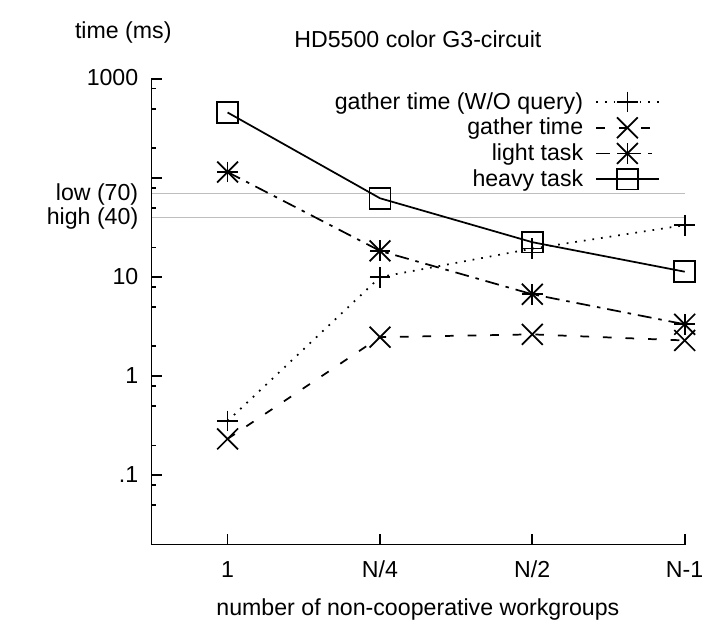} \\
\includegraphics[width=.7\columnwidth]{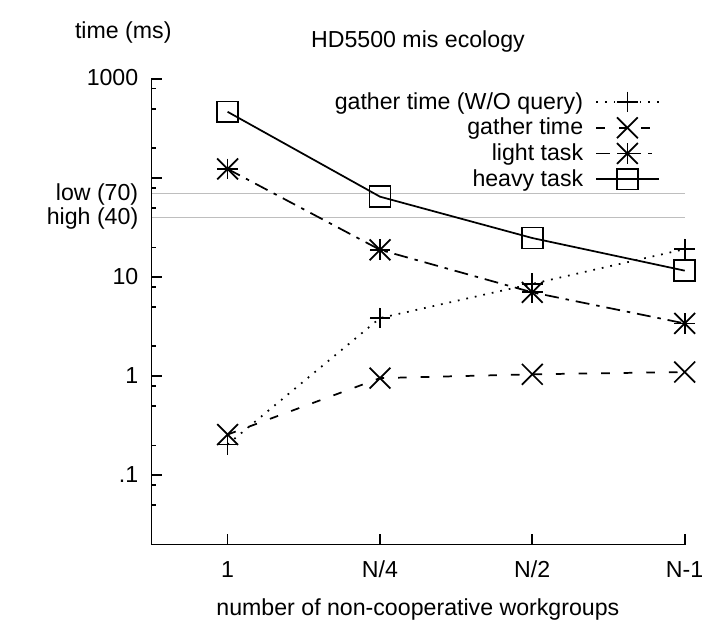} \\
\includegraphics[width=.7\columnwidth]{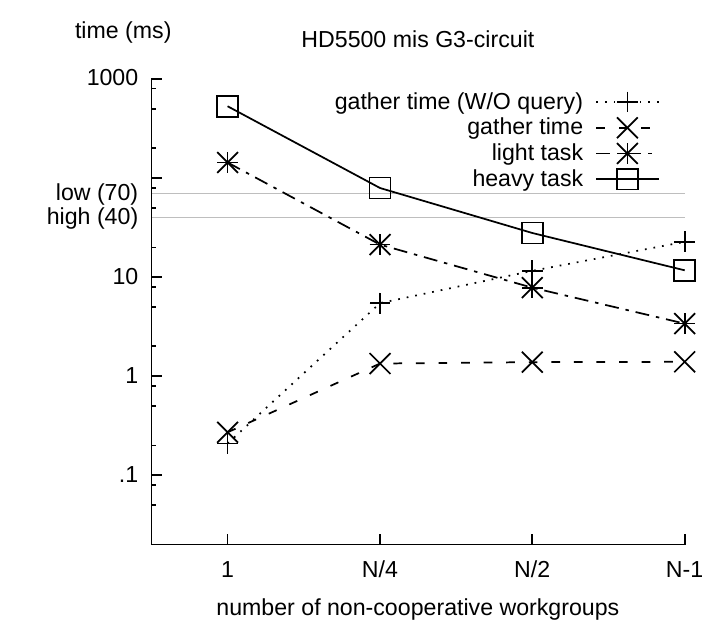} \\
\includegraphics[width=.7\columnwidth]{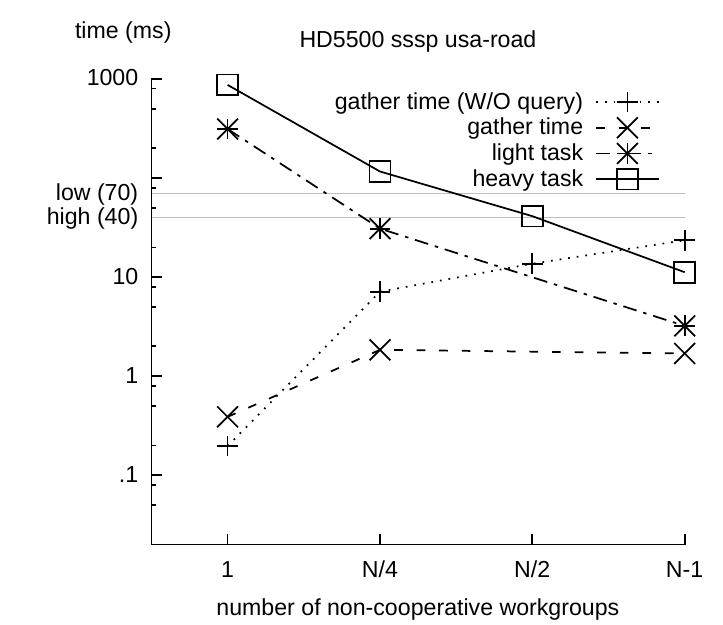} \\
\includegraphics[width=.7\columnwidth]{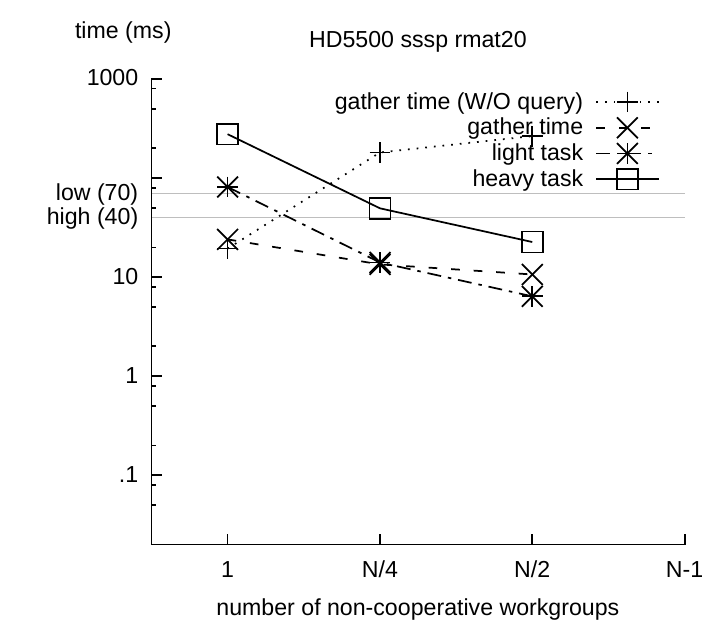} \\
\includegraphics[width=.7\columnwidth]{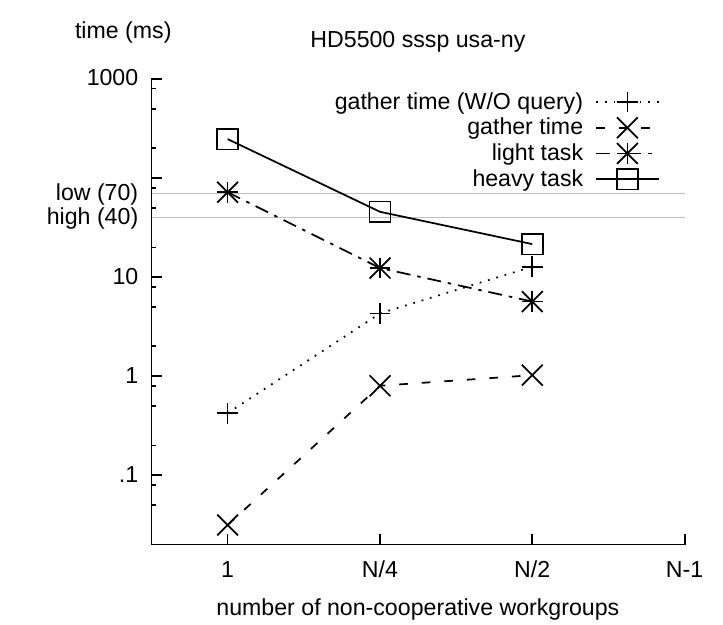} \\
\includegraphics[width=.7\columnwidth]{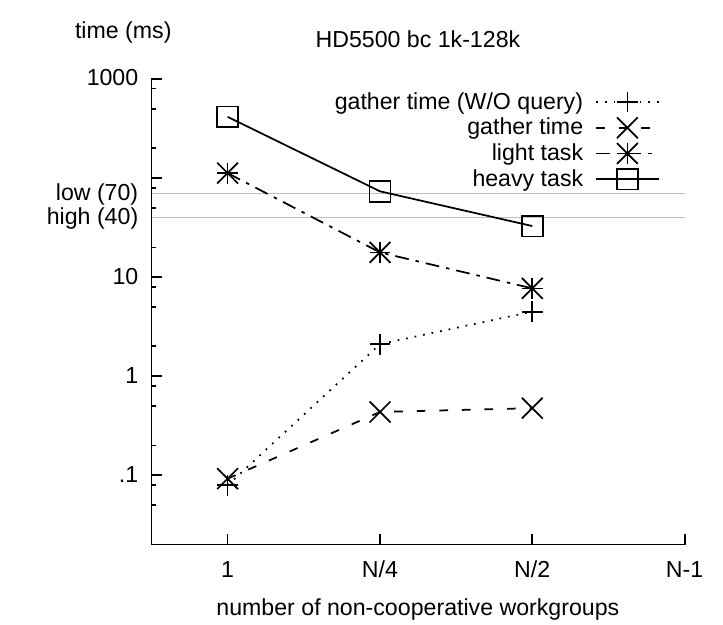} \\
\includegraphics[width=.7\columnwidth]{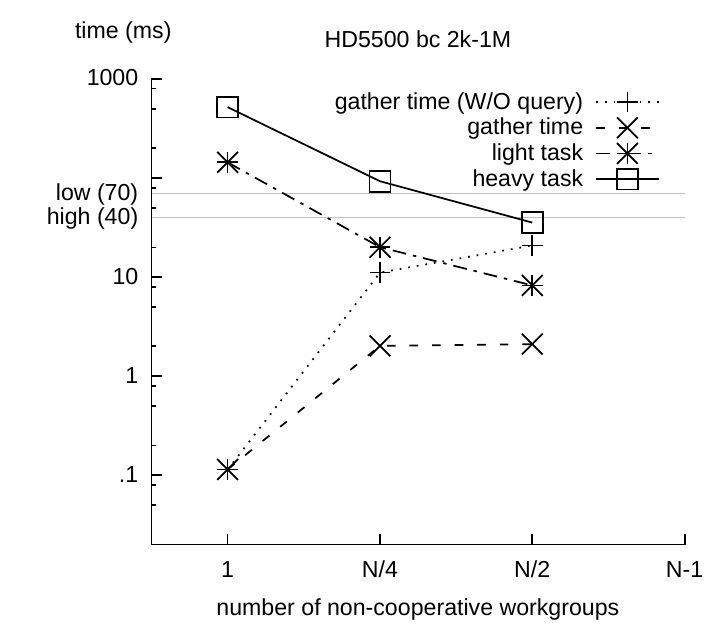} \\
\includegraphics[width=.7\columnwidth]{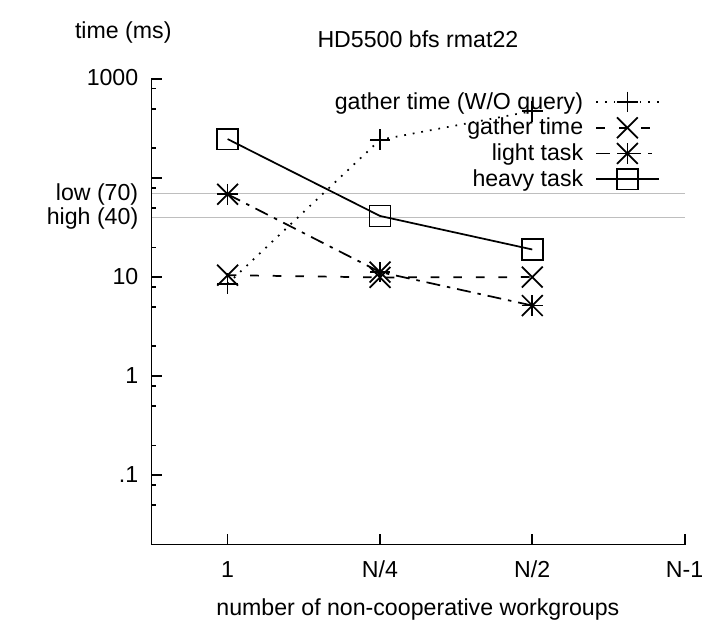} \\
\includegraphics[width=.7\columnwidth]{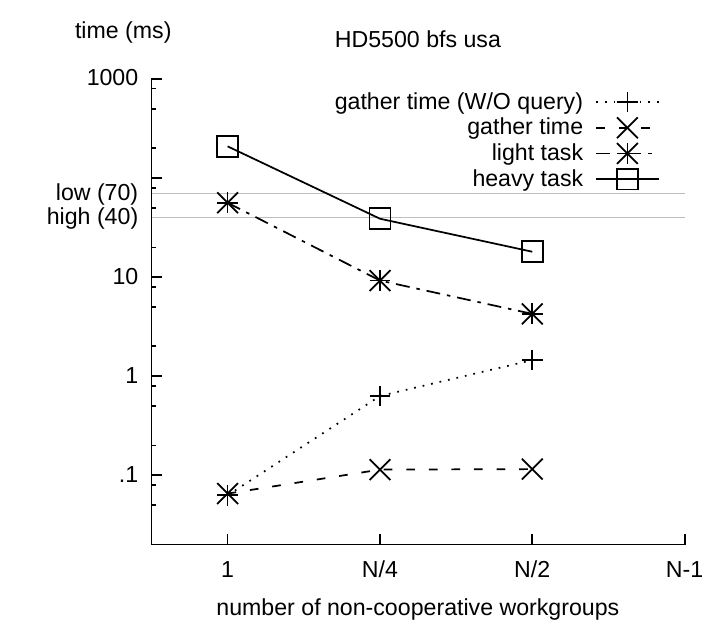} \\
\includegraphics[width=.7\columnwidth]{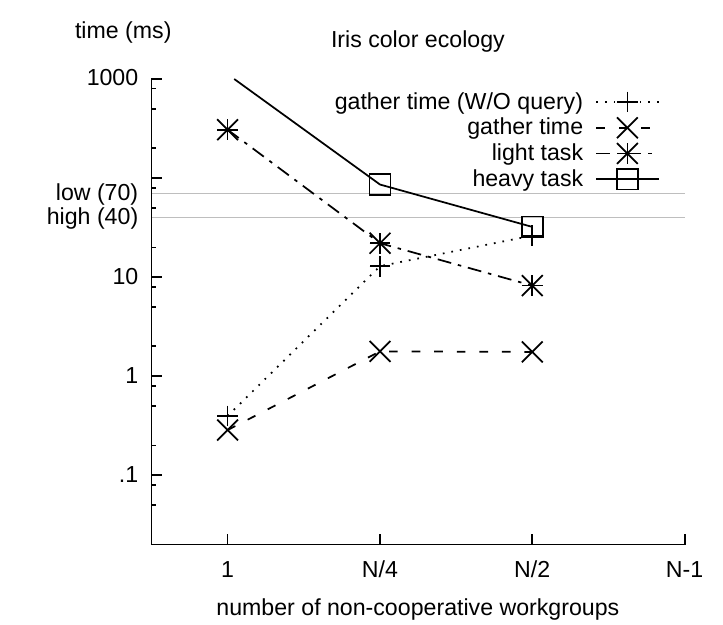} \\
\includegraphics[width=.7\columnwidth]{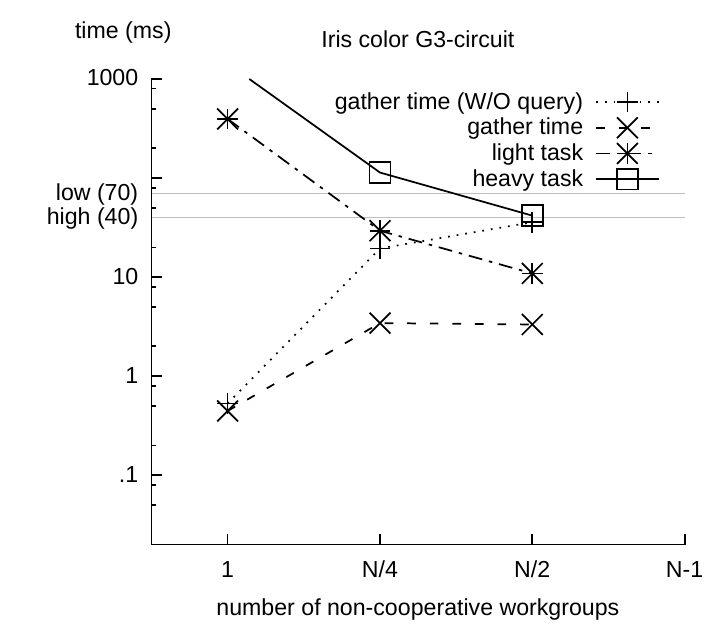} \\
\includegraphics[width=.7\columnwidth]{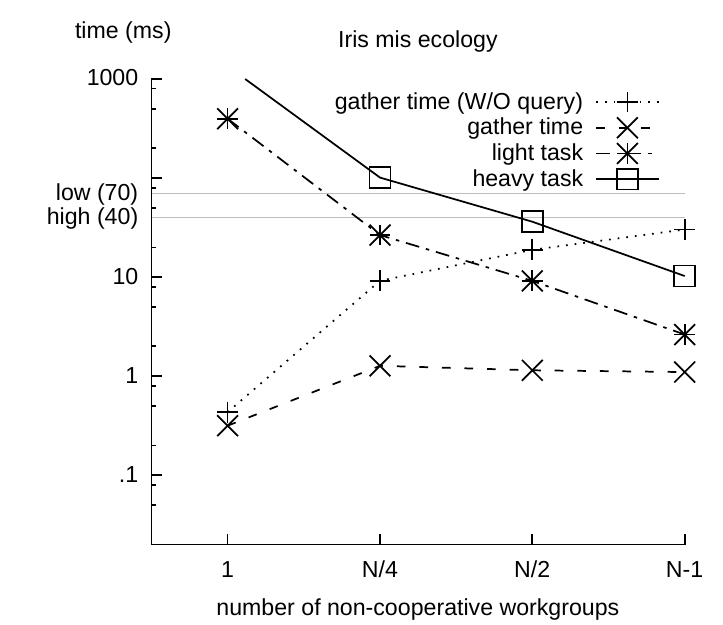} \\
\includegraphics[width=.7\columnwidth]{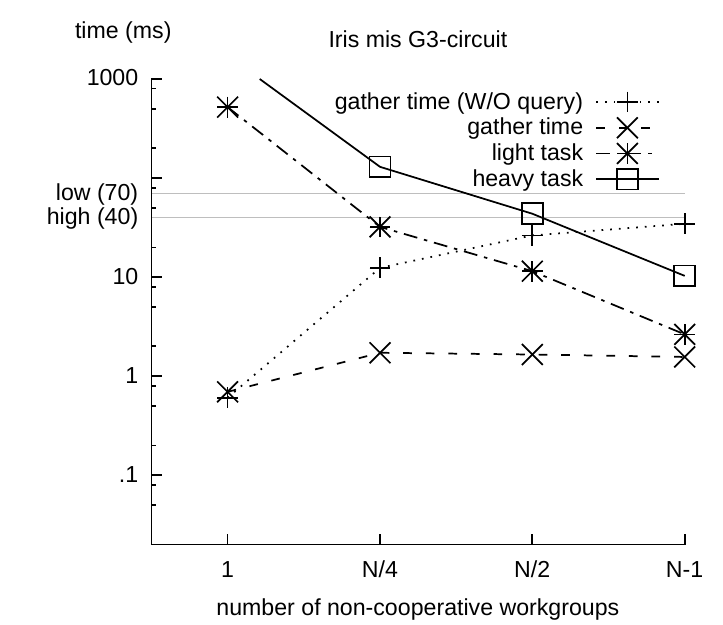} \\
\includegraphics[width=.7\columnwidth]{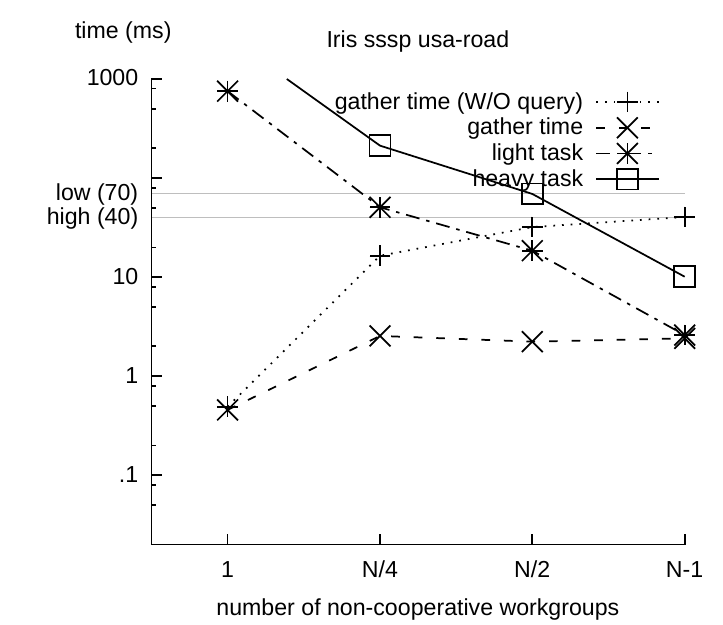} \\
\includegraphics[width=.7\columnwidth]{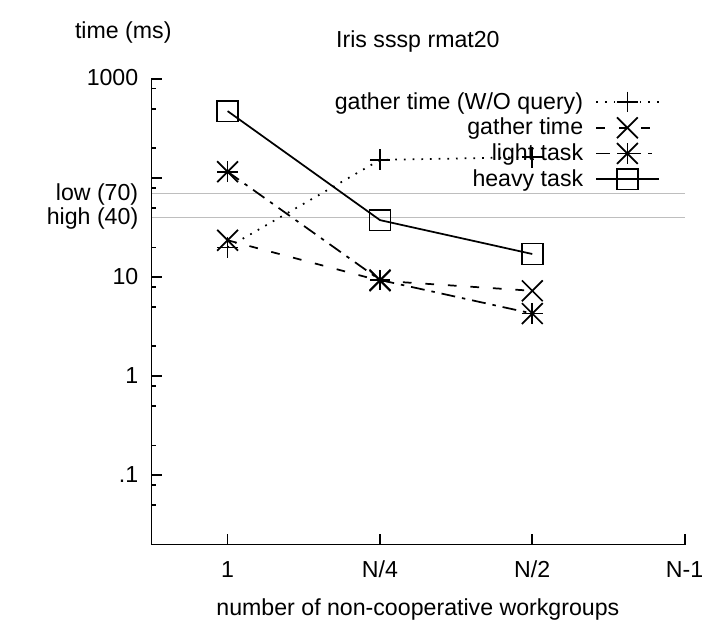} \\
\includegraphics[width=.7\columnwidth]{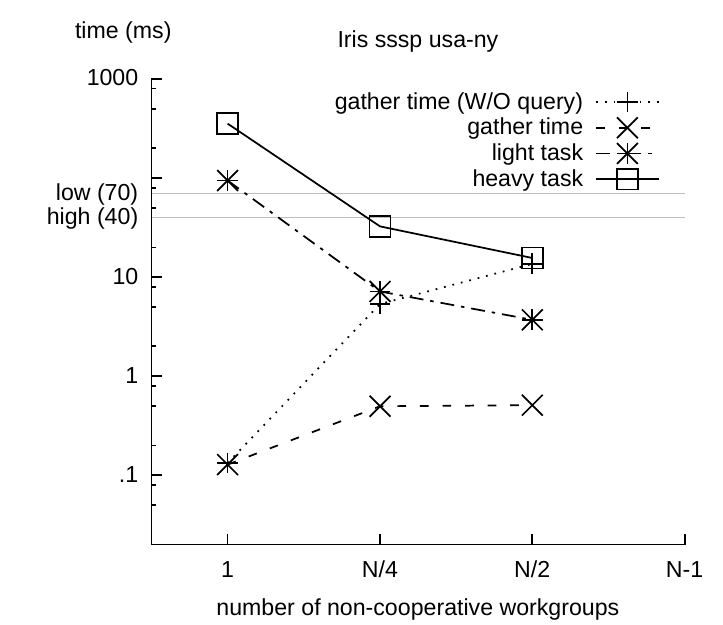} \\
\includegraphics[width=.7\columnwidth]{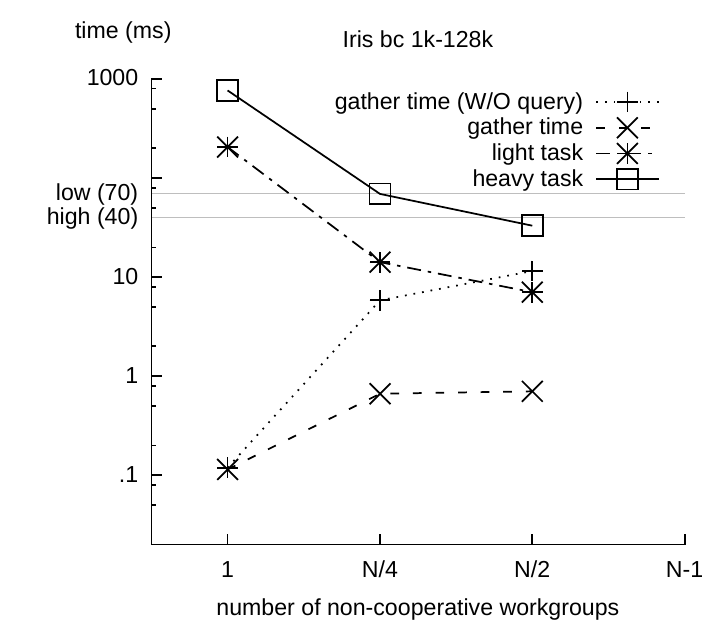} \\
\includegraphics[width=.7\columnwidth]{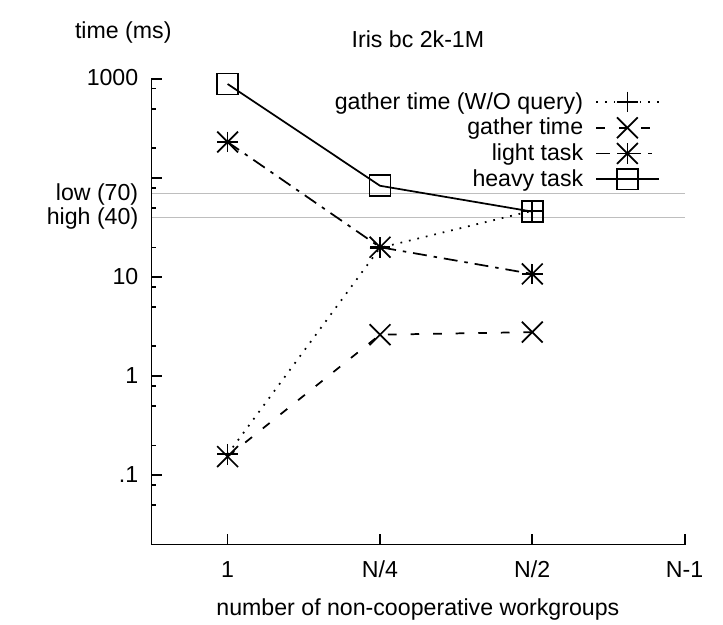} \\
\includegraphics[width=.7\columnwidth]{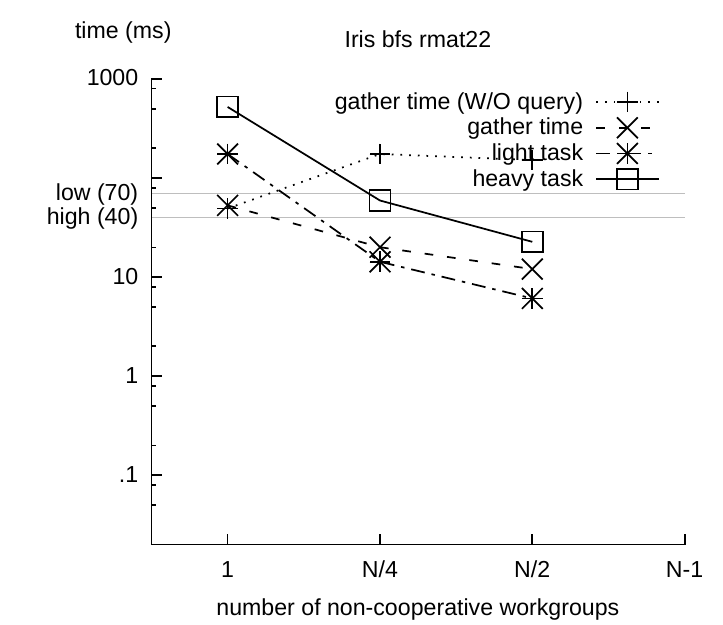} \\
\includegraphics[width=.7\columnwidth]{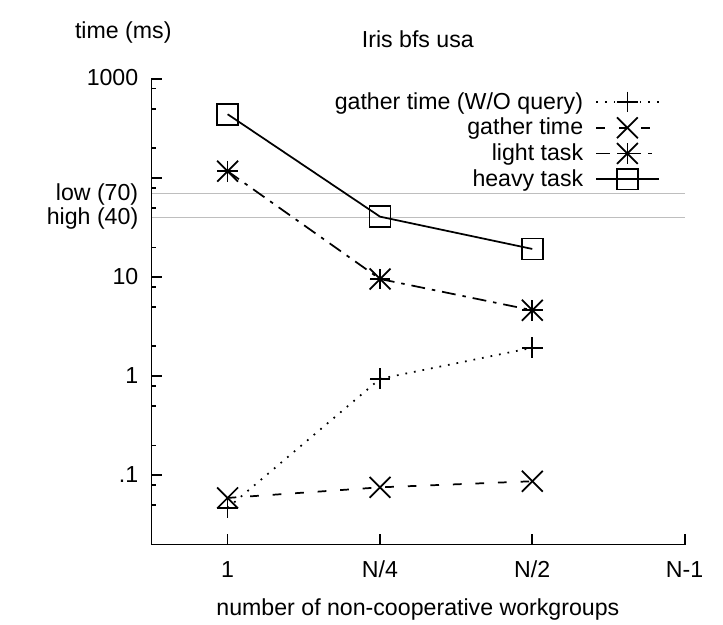} \\
\includegraphics[width=.7\columnwidth]{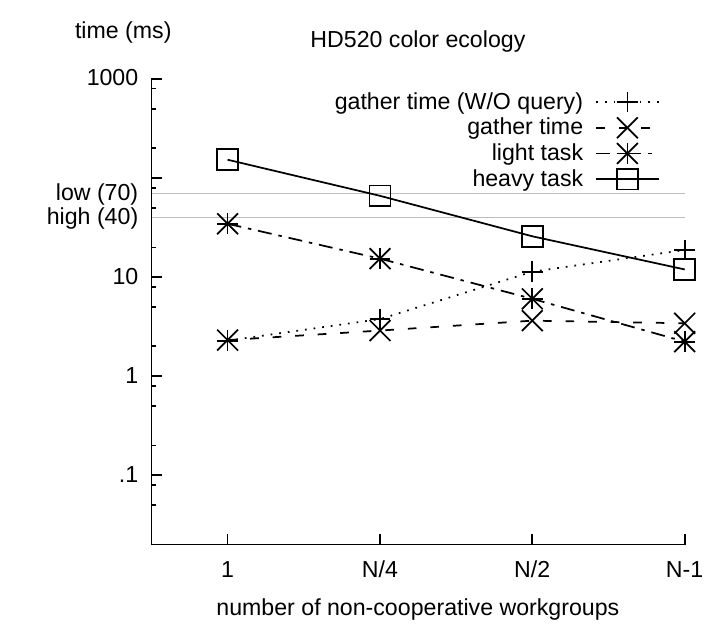} \\
\includegraphics[width=.7\columnwidth]{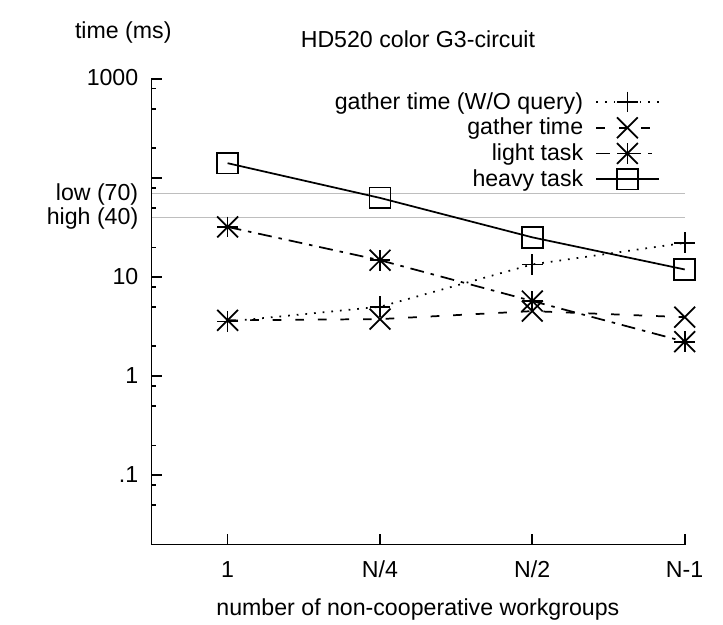} \\
\includegraphics[width=.7\columnwidth]{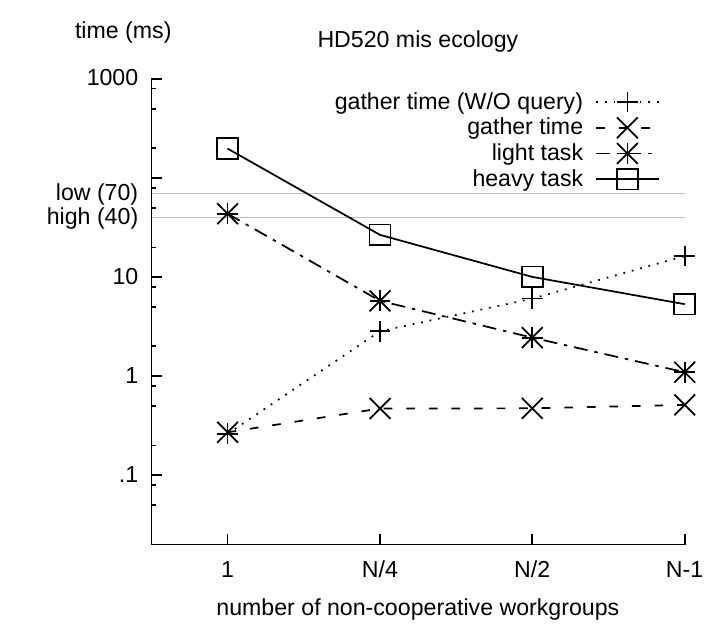} \\
\includegraphics[width=.7\columnwidth]{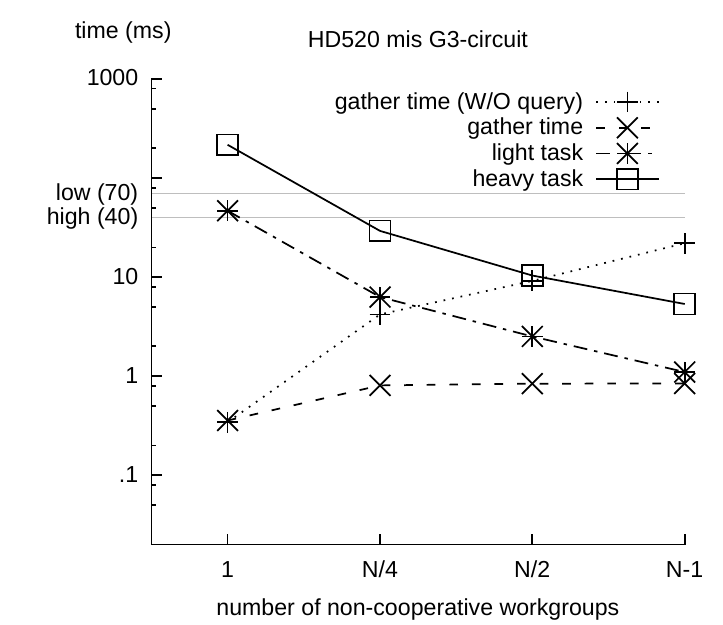} \\
\includegraphics[width=.7\columnwidth]{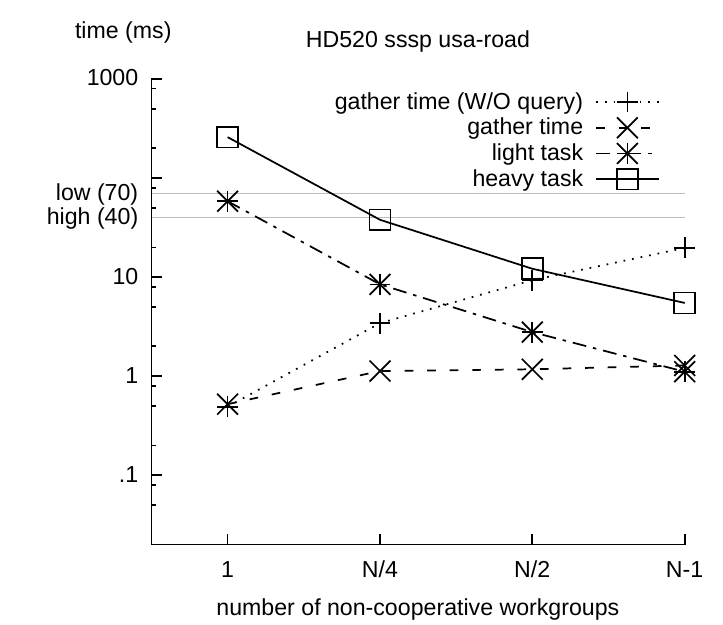} \\
\includegraphics[width=.7\columnwidth]{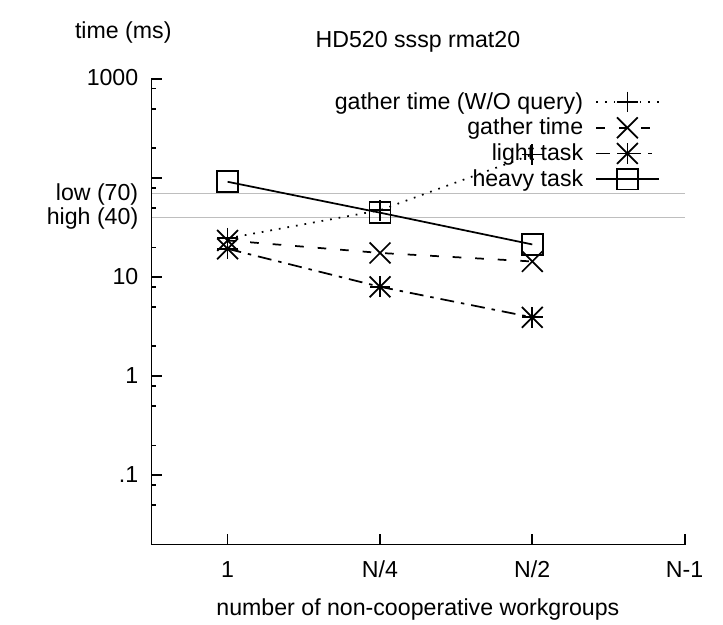} \\
\includegraphics[width=.7\columnwidth]{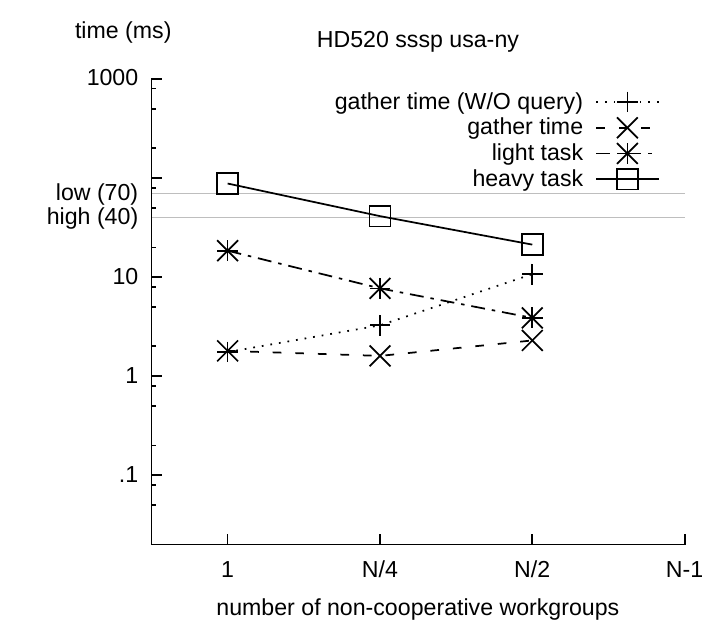} \\
\includegraphics[width=.7\columnwidth]{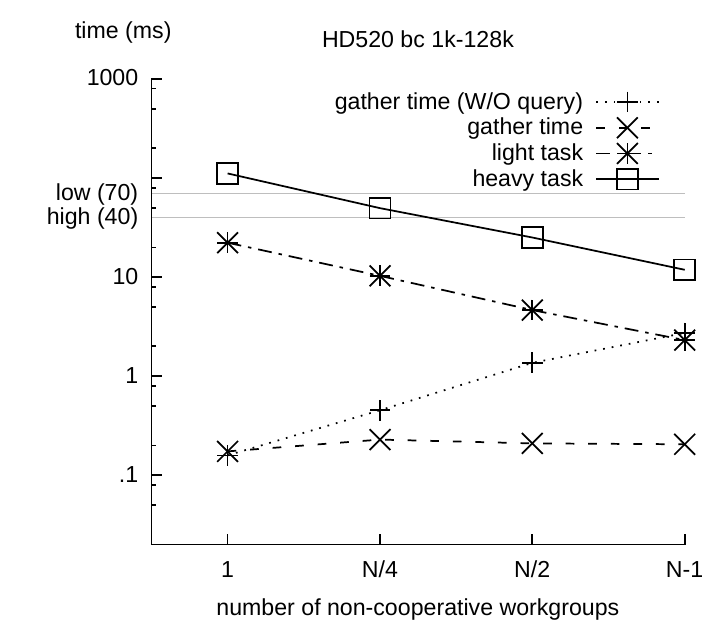} \\
\includegraphics[width=.7\columnwidth]{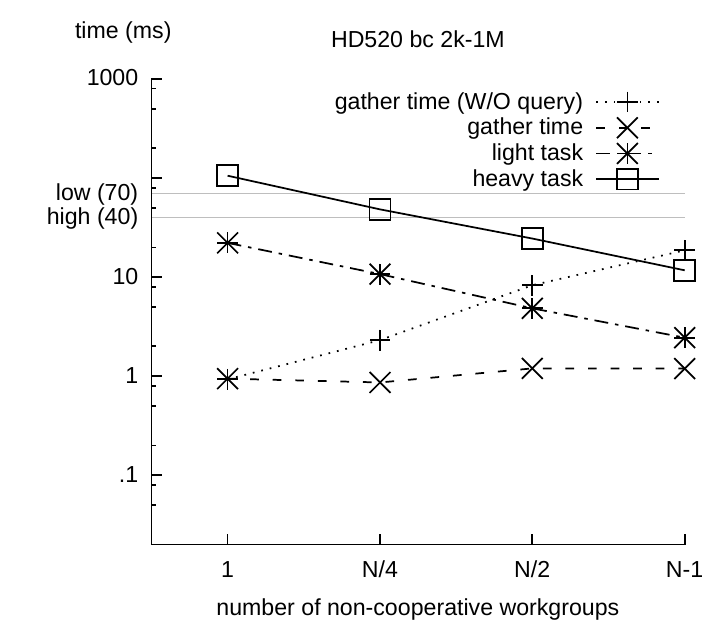} \\
\includegraphics[width=.7\columnwidth]{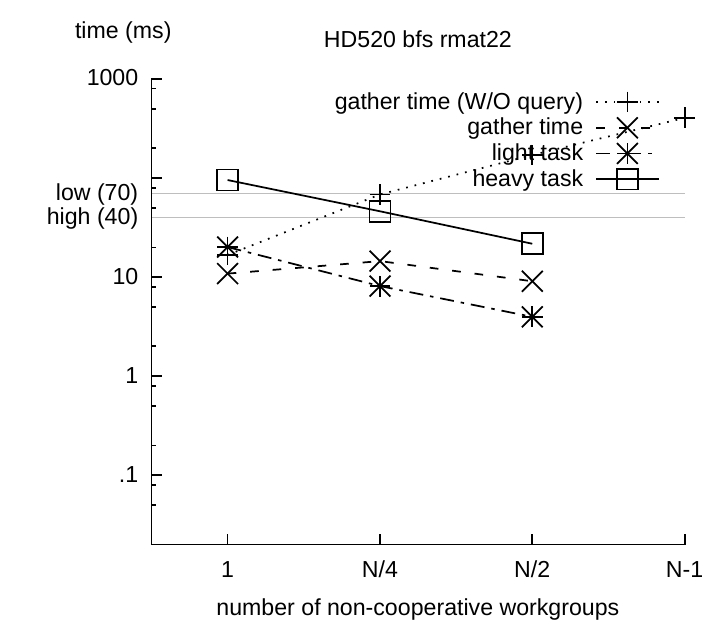} \\
\includegraphics[width=.7\columnwidth]{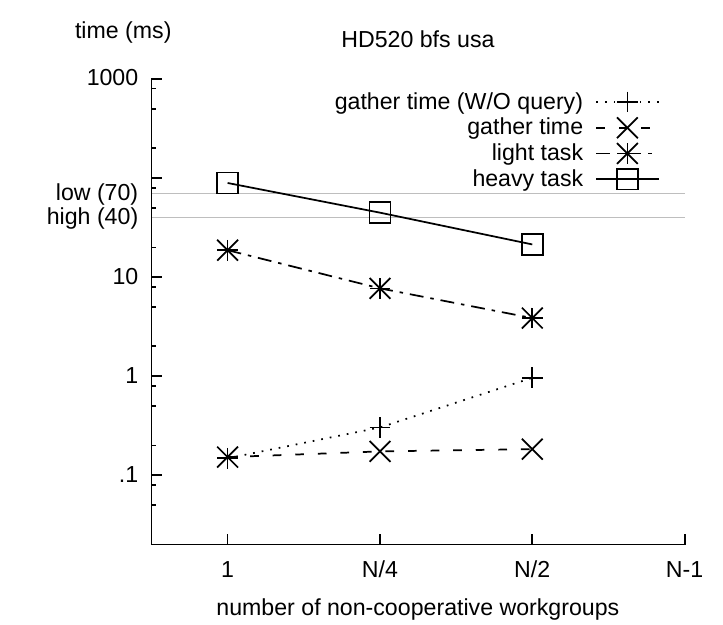} \\
\includegraphics[width=.7\columnwidth]{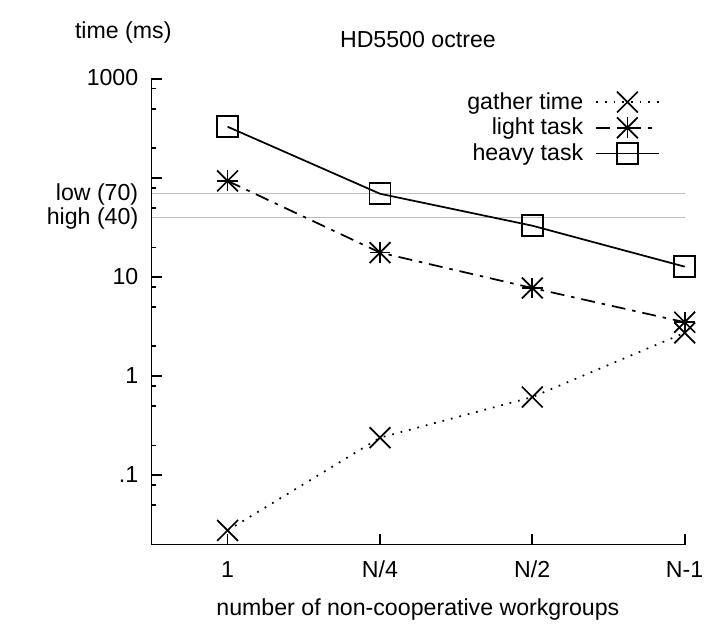} \\
\includegraphics[width=.7\columnwidth]{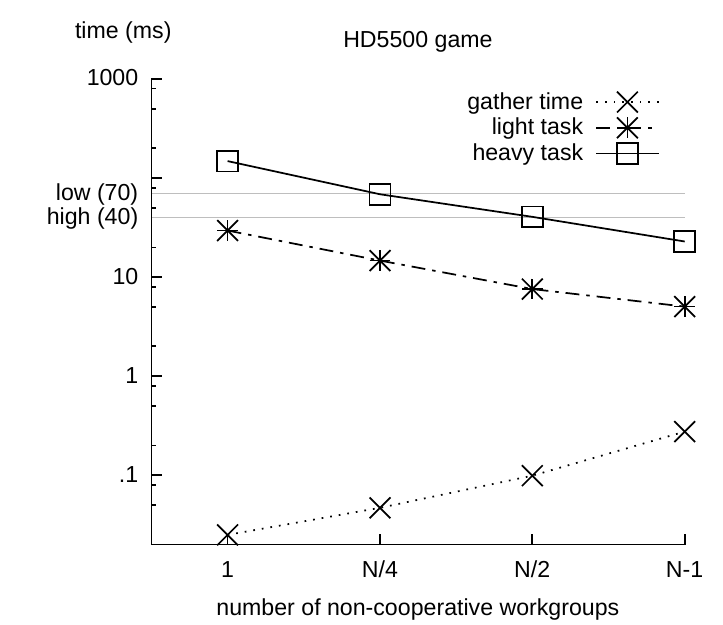} \\
\includegraphics[width=.7\columnwidth]{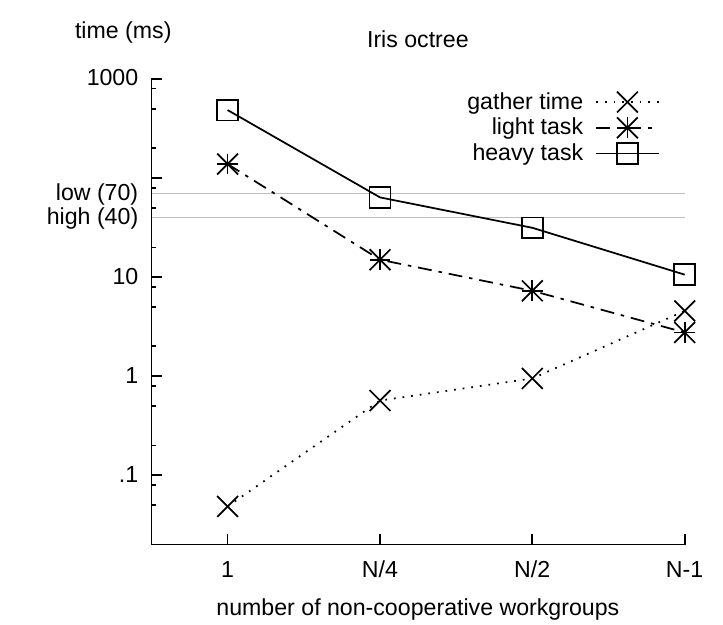} \\
\includegraphics[width=.7\columnwidth]{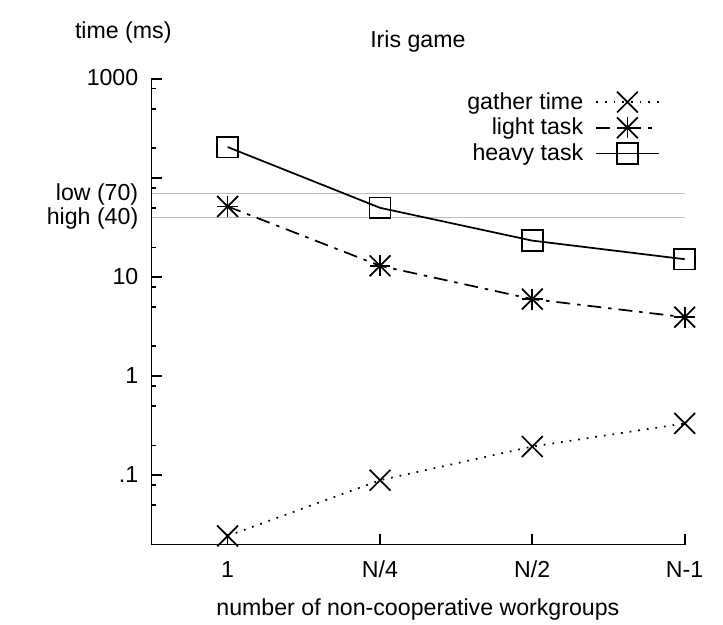} \\
\includegraphics[width=.7\columnwidth]{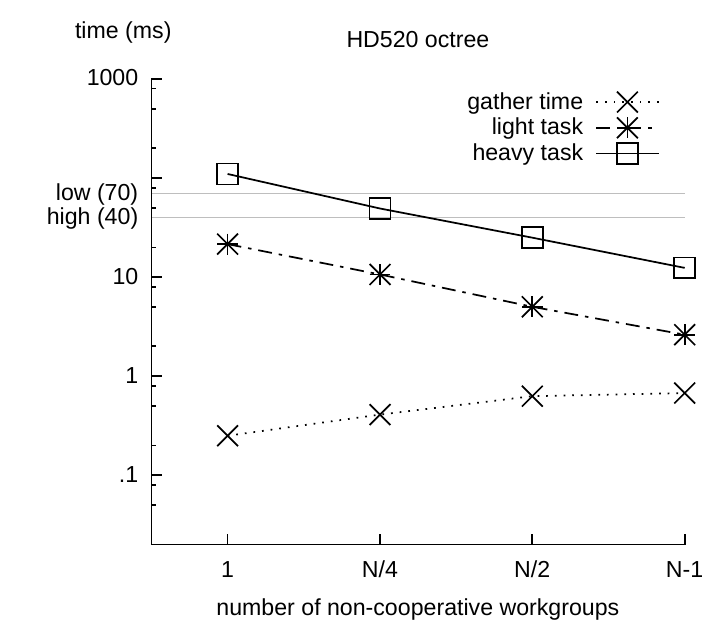} \\

\end{document}